%% file: task_stealing_exahype_arxiv.tex
\documentclass{article}
\usepackage{amsmath,amssymb,amsfonts}
\usepackage{amsthm}
\usepackage{graphicx}
\usepackage{textcomp}
\usepackage{xcolor}
\usepackage{subcaption}
\usepackage{algorithm}
\usepackage{algpseudocode}
\usepackage{hyperref}

\usepackage{comment}
\usepackage{authblk}

\newtheorem{definition}{Definition}
 
\usepackage[final]{changes}
\definechangesauthor[name=CCPE/R1, color=orange]{CCPE1}

\title{Lightweight Task Offloading Exploiting MPI Wait Times for Parallel Adaptive Mesh Refinement}

\author[1]{Philipp Samfass\thanks{samfass@in.tum.de}}

\author[2]{Tobias Weinzierl}

\author[2]{Dominic E. Charrier}

\author[1]{Michael Bader}

\affil[1]{Department of Informatics\\ Technical University of Munich, Garching, Germany}
\affil[2]{Computer Science\\ Durham University, Durham, Great Britain}

\begin{document}
\maketitle

\begin{abstract}
\input{00_abstract}
\end{abstract}

\input{01_introduction}

\input{02_exahype}
\input{03_vision.tex}

\input{04_terminology}

\input{05_algorithm}

\input{07_implementation}

\input{08_results}

\input{09a_discussion}

\input{09b_conclusion}

\input{99_acknowledgements}

\bibliographystyle{plain}
\bibliography{task_stealing_exahype}  

\end{document}

%% file: 00_abstract.tex
Balancing \replaced[id=CCPE1]{the workload of sophisticated
simulations}{dynamically adaptive mesh refinement (AMR) codes} 
is inherently difficult, since \replaced[id=CCPE1]{we}{codes} have to
balance both computational workload and memory footprint over meshes that can change any time
\added[id=CCPE1]{or yield unpredictable cost per mesh entity}, while modern supercomputers
and their interconnects start to exhibit fluctuating performance.
We propose a novel lightweight \replaced[id=CCPE1]{balancing technique}{scheme}
for MPI+X \replaced[id=CCPE1]{to accompany}{which complements} traditional\added[id=CCPE1]{, prediction-based load} balancing. 
It is a reactive diffusion approach \replaced[id=CCPE1]{that}{which} uses online
measurements of MPI idle time to migrate tasks
\replaced[id=CCPE1]{\emph{temporarily}}{non-persistently}
from overloaded to underemployed ranks. Tasks are deployed to ranks which otherwise would wait, processed with high priority, and made available to the overloaded ranks again. \replaced[id=CCPE1]{This migration is non-persistent.}{They are temporarily migrated.} 
Our approach hijacks idle time to do meaningful work and is totally non-blocking, asynchronous and distributed without a global data view.
Tests with a seismic simulation code \deleted[id=CCPE1]{running an explicit high
order ADER-DG scheme (}developed in the ExaHyPE engine\deleted[id=CCPE1]{,
www.exahype.org)} uncover the method's potential. We found speed-ups of up to
2--3 for ill-balanced scenarios without logical modifications of the
code base \added[id=CCPE1]{and show that the strategy is capable to react quickly to temporarily changing
workload or node performance.
}

%% file: 01_introduction.tex

\section{Introduction}
\label{section:introduction}

%
%
Load balancing that decomposes work prior to
a certain compute phase---a time step or iteration of an equation system
solver---is doomed to underperform in many sophisticated simulation codes.
There are multiple reasons for this:
The clock frequency of processors changes over runtime~\cite{Charrier:19:EnergyAndDeepMemory,Acun:16:Variation,Charles:09},
the network speed is subject to noise due to other applications~\cite{Jain:2017,Pollard:2018} 
or IO, and
task-based multicore parallelisation (MPI+X) tends to yield fluttering
throughput due to effects of the memory hierarchy \cite{McCalpin:2018}, work
stealing and non-determinism in the MPI progression.
While this list is not comprehensive, notably modern numerics drive the
non-predictability:
They build atop of dynamic adaptive mesh refinement (AMR)
that changes the mesh throughout a time step or mesh sweep
\added[id=CCPE1]{\cite{Charrier:19:EnclaveTasking}}, combine different physical
models
\added[id=CCPE1]{\cite{Charrier:19:EnclaveTasking,Dumbser:2006,Jofre:15:ParallelLB}}, or solve non-linear equation systems with iterative solvers in substeps
\added[id=CCPE1]{\cite{Reinarz:2019}}.
It becomes hard or even impossible to predict a step's computational load.
As adjusting parallel partitions and respective data
migration is often costly, many AMR codes consequently repartition only every
10th or 100th time step and tolerate certain load imbalances
\added[id=CCPE1]{in-between}.

%
%
We propose a novel, lightweight load redistribution scheme
that acts on top of traditional load balancing.
It\added[id=CCPE1]{, firstly,}  assumes that parts of the underlying simulation
code are phrased in terms of \added[id=CCPE1]{many expensive}
tasks. 
\added[id=CCPE1]{
 It, secondly, assumes that good AMR codes manage to hide data exchange
 behind computations yet can not keep all cores busy all the time.  
 In every solver step, some cores on some ranks have to wait for MPI data to
 drop in.
} 
Our idea is to offload tasks from overbooked to waiting ranks to make these
work productively rather than being idle
\added[id=CCPE1]{\cite{Garcia:09:LeWI}}.
The code plugs into the MPI operations searching
for the late sender pattern \cite{Mao:14:WaitStateProfiler},
which yields a wait graph. 
Ranks \replaced[id=CCPE1]{that find out that}{identify that} they are critical
to the walltime\deleted[id=CCPE1]{,} search for ``optimal victims'', i.e.~ranks
that can take up further work without slowing down the overall computation,
and then actively offload tasks to victim ranks.
%
%
\added[id=CCPE1]{Thirdly, }
we assume that neither load distribution nor imbalances change
radically in-between algorithm steps.
We therefore update the wait graph on-the-fly, 
using concepts from reinforcement learning \cite{Bottou:18:optimization},
and let the wait graph guide a diffusion 
of tasks to follow load alterations.
\added[id=CCPE1]{
Finally, we keep local copies of all offloaded tasks.
This allows us to urgently recompute them if the temporary outsourcing does
not come back with results fast enough.
We overaggressively distribute tasks to build up a load balancing slack, and 
thus can react quickly to unforeseen load imbalances.
}

%
%
Task-based parallelisation between MPI ranks is not new.
The UIntah framework~\cite{Davison:00:Uintah,Meng:2010:Uintah} for example uses a centralised
data/task warehouse from which ranks are served.
Tasks therefore are not tied to a particular rank and the ownership is
(logically) with the warehouse.
The Swift project~\cite{Schaller:16:Swift} as another example phrases a whole SPH simulation in
terms of tasks and applies graph partitioning to derive task
decomposition and task migration patterns over the whole machine, i.e.~both
shared and distributed memory domains.
This is a wholistic, fine-granular, proactive load balancing approach.
Charm++~\cite{Acun:2014:Charm} features tasks that can be migrated and a runtime
which tracks task dependencies in-between ranks dynamically.
Dependencies thus pose no constraint on the task placement.
Other task-based approaches such as HPX~\cite{Kaiser:14:HPX,Software:HPX} feature task migration between different
processes via a global address space.
\replaced[id=CCPE1]{%
 The AMR framework sam(oa)$^2$ finally introduces task stealing driven by 
 the application
 \cite{Samfass:18:ReactiveSamoa} in-between bulk-synchronous processing.
 This list is not comprehensive.
}
{The AMR framework sam(oa)$^2$ finally demonstrates a reactive task stealing
approach that is entirely driven by the application
\cite{Samfass:18:ReactiveSamoa} in-between bulk-synchronous processing.
}

%
%
\added[id=CCPE1]{
An established alternative to a functional decomposition---typically realised
through tasks---is a data,
i.e.~domain decomposition.
In an AMR world or situations where the load per cell is hard to 
predict, it has
to be combined with frequent rebalancing.
Efficacious load re-balancing strategies relying on space-filling 
curve cuts,
diffusion processes or graph algorithms for example are known.
Several properties determine whether they yield effective, i.e.~fast, code:
First, an appropriate geometric cost model has to exist.
If energy constraints compromise the compute nodes' performance 
\cite{Charrier:19:EnergyAndDeepMemory},
if numerical schemes yield unpredictable workload per mesh entity, or if 
different physical
models are applied to the same mesh set, deriving a cost model becomes
non-trivial.
Second, memory constrains the balancing.
If very cheap and very expensive grid areas coexist, 
situations can
arise where a load balancer can not fit a big enough (cheap) 
subpartition to one resource.
Third, data transfer cost constrains rebalancing.
Even once a good domain decomposition is determined, the cost of moving 
towards
this good decomposition from a given partitioning can outweight the 
gain if the partitioning remains advantageous only for few compute steps.
Finally, spatial redistribution is an algorithmic step which synchronises
resources and stresses the communication subsystem.
If many nodes rebalance at the same time, the communication subsystem is
heavily used though there might have been periods of underutilisation 
throughout compute phases.
}

%
%
\replaced[id=CCPE1]{
 While our approach starts from existing load balancing and takes up 
 ideas and extends upon existing work,
 it introduces new capabilities:%
}
{While our approach takes up ideas and extends upon existing work, 
it differs in the following fundamental ways:}
(i) It does not target load distribution per se but determines MPI waiting times to improve
upon existing load balancing. 
\replaced[id=CCPE1]{This improvement}{It} is a reactive rather than a
predictive add-on to load balancing
\added[id=CCPE1]{and notably does not require an a priori cost model
\cite{Jofre:15:ParallelLB}.} 
(ii) It is very fine-grained as it acts on the level of individual 
(compute-intense)
tasks.
Yet, no task dependencies are tracked.
We work non-persistently. Tasks are offloaded to other ranks, processed
there, and the results are immediately sent back. 
(iii) It is a lightweight approach since the task migration is realised through
a set of tasks itself. 
Therefore, we plug seamlessly into the tasking system and the overhead is
small.
We do not need a dedicated load balancing or MPI progression thread
\cite{Hoefler:08:SacrificeThread}.
\deleted[id=CCPE1]{
(iv) It gives ranks the opportunity to outsource tasks. This opportunity
``window'' is opened reactively, i.e.~ranks know how many tasks they are allowed
to give away and they consequently use this for non-urgent tasks:
Tasks are given away proactively, i.e.~prior to
one rank running idle.
We bring together the best of two worlds: proactive task outsourcing and
the reaction to runtime changes.
This strategy differs from the aforementioned approaches.
}%
To our knowledge, this is
the first approach that abandons the attempt to perfectly balance work in a predictive way but rather explicitly determines and hijacks MPI wait times to guide
a lightweight task distribution
\replaced[id=CCPE1]{while it remains reactive without the latency penalty
introduced by classic task stealing, i.e.~it can react to quickly changing
performance and load balancing characteristics.}{.}

%
%
Its properties render our approach promising for many applications
which are already phrased in tasks. 
We assess it by means of an earthquake simulation benchmark.
The underlying code base
ExaHyPE \cite{Software:ExaHyPE,Reinarz:2019}
relies on an explicit time stepping scheme, which works on dynamically adaptive
meshes.
The setup poses a challenge to our approach as it is not dominated
by few compute-intense tasks. 

We benchmark the reactive scheme against sole geometric domain
decomposition and against a task distribution which is derived from 
chains-on-chains partitioning (CCP) \cite{Pinar:04:CCP}.

%
%
Our manuscript is organised as follows: 
We introduce the benchmark code in Sect.~\ref{section:exahype} before we phrase
our vision (Sect.~\ref{section:vision}).
Some terminology (Sect.~\ref{section:terminology}) 
allows us to introduce a set of load balancing strategies in
Sect.~\ref{section:load-balancing}. 
This core contribution starts from a point-to-point diffusion approach which is
augmented and accelerated by various techniques. 
In Sect.~\ref{section:implementation}, we
elaborate on the technical details of our implementation.
Some experiments in Sect.~\ref{section:results} highlight the potential of the
approach.
We close the discussion with an interpretation of the 
scheme's characteristics (Sect.~\ref{section:discussion}), before we identify
further application areas of the proposed methodology plus future work in
Sect.~\ref{section:outlook}.

%% file: 02_exahype.tex

\section{A parallel ADER-DG seismic solver on adaptive meshes }
\label{section:exahype}

%
%
Our benchmark code implements an explicit high order discontinuous Galerkin solver 
for the linear elastic wave equations, which may be written as (cf.\ \cite{igel2016}, e.g.)
\begin{eqnarray*}
  \frac{\delta\sigma}{\delta t} - E(\lambda,\mu) \cdot \nabla v  &= & S_\sigma,
  \\
  \frac{\delta v}{\delta t} - \frac{1}{\rho} \nabla \cdot \sigma &= & S_v,
  \label{eq:ElasticWaveEquation_Newton}
\end{eqnarray*}

\noindent
with a velocity field $v$ and a stress tensor $\sigma $ in the first equation of
the system.
It results from Hooke's law and evolves both quantities through a stiffness
tensor $E$ depending on the Lam\'e constants $\lambda$ and $\mu$ (i.e., material parameters).
The second equation describes Newton's second law.
$\rho$ here is the density of the material.

As simulation setup, we use the established Layer Over Halfspace~1 (LOH.1) benchmark \cite{Day:loh1}.
It mimics an earthquake via a simplified setting that assumes a point source in a cubic domain 
that consists of two material layers:
a thin sediment layer (with slower wave speeds) over a rock layer (with higher wave speeds).
LOH.1 is part of a widely-used collection of benchmark scenarios to validate codes 
and compare results with other simulation software.

\subsection*{ADER-DG: High Order Discontinuous Galerkin}

Our solver realises an Arbitrary high-order DERivative
\replaced[id=CCPE1]{Discontinuous}{discontinuous} Galerkin (ADER-DG) method
\cite{zanotti:2015} on \replaced[id=CCPE1]{tree-structured}{tree-structure}
Cartesian grids.
It is implemented within the ExaHyPE engine to solve hyperbolic PDE systems \cite{Reinarz:2019}.
In the following, we summarize the main computational steps of the scheme,
whereas we describe full details of the scheme in previous work \cite{Charrier:19:AderDG}.

ADER-DG is an explicit time-stepping scheme that decomposes each time step into three phases, 
thus computing $(\sigma ,v)(t+\Delta T) = (\mathcal{C} \circ \mathcal{R} \circ \mathcal{P})(\sigma , v)(t)$. 
Each grid cell approximates the solution locally via a tensor product of polynomials of degree $p$
(orthogonal polynomials constructed on Gauss-Legendre points), 
following a classic DG-SEM (DG Spectral Element Method) approach \cite{hesthaven:nodalDG}.
In the \emph{predictor} step $\mathcal {P}$,
the algorithm first extrapolates the solution in time, 
ignoring the influence of neighbouring cells and evolves $ (\sigma , v) $.
This step follows the Cauchy-Kovaleskaya procedure \cite{Dumbser:2006}.
The arising discontinuities in the predicted solution  $ (\sigma , v) $ along the cell faces are
next subject to a space-time Riemann solver $\mathcal{R}$.
Finally, we bring the Riemann solution and the predicted value
together, i.e.,~correct ($\mathcal{C}$) the predicted value.

Our code discretises our computational domain through a spacetree
 \cite{Weinzierl:19:Peano} 
and thus solves the problem on an adaptive Cartesian
grid where the individual cells are cubes.
Each cell may be transformed according to a curvi-linear transformation\replaced[id=CCPE1]{\cite{hesthaven:nodalDG}}{(as in \cite{hesthaven:nodalDG})} to align to geometry features: 
Each cell carries a transformation matrix which fits it to the actual topology, 
allowing simulation of seismic wave propagation in complex topographies. 
For the LOH.1 benchmark, the transformation matrix is simple (but causes the same 
computational load), as it only aligns the material discontinuity in the LOH.1 
geometry to our Cartesian grid. 
\replaced[id=CCPE1]{To reduce the discretisation error further, we}{We}
adaptively  refine the mesh in the top sediment layer and around the point source.
\deleted[id=CCPE1]{We coarsen towards the other boundaries to reduce reflections
due to non-perfectly absorbing boundary conditions.}

\begin{figure}
  \begin{center}
    \includegraphics[width=0.4\textwidth]{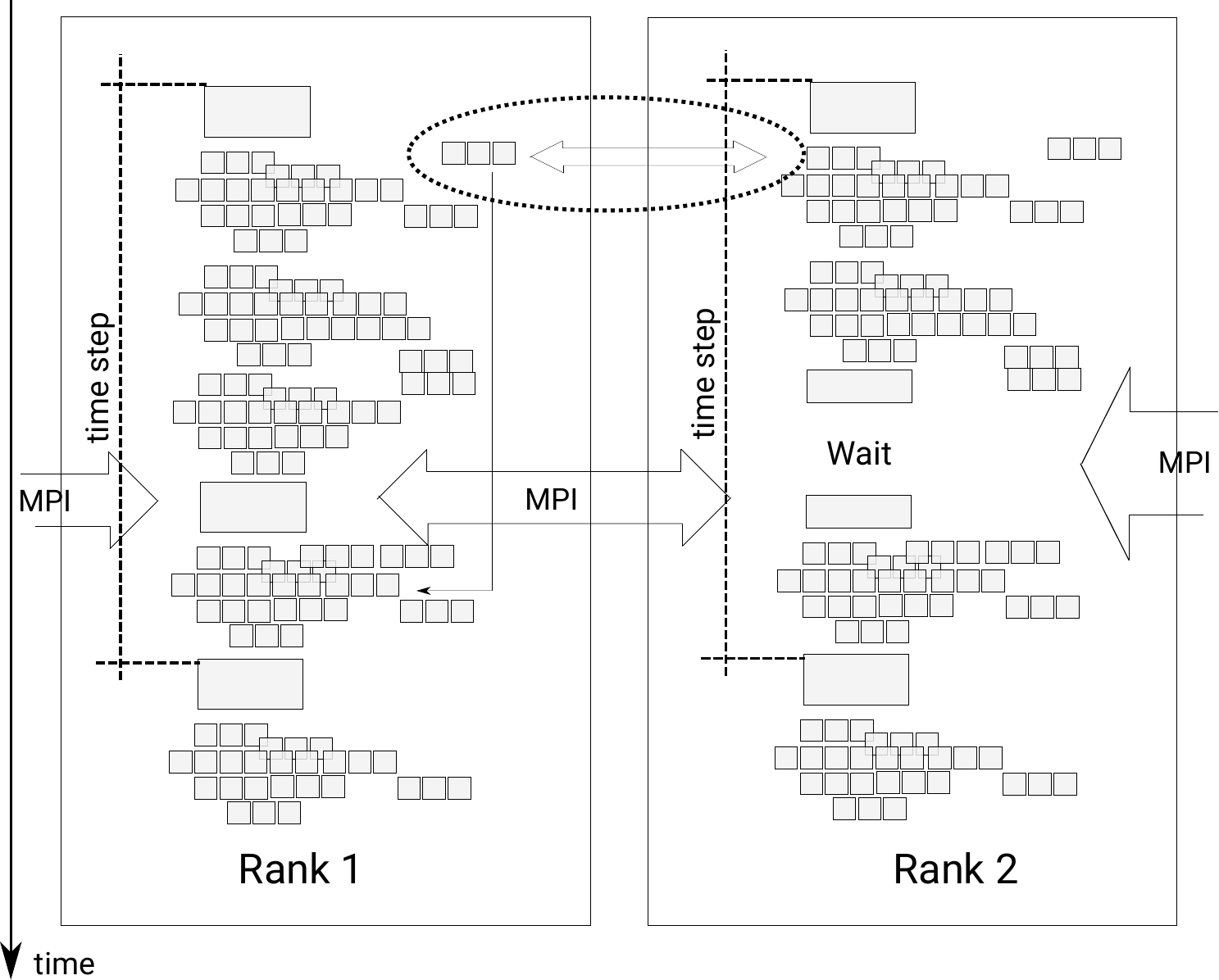}
  \end{center}
  \caption{
      Schematic program execution:
      Ranks decompose into tasks. Some tasks' outcomes are required only late 
      throughout the computation or even after the boundary data
      exchange or in the next time step.
      High bandwidth demands arise towards the end of the time step, i.e.,~we are
      not consistently bandwidth-bound.
      Our scheme offloads non-urgent tasks to MPI ranks that tend to wait and
      immediately transfer the outcome back (tasks within the dotted circle belong
      to rank 1 but are computed on rank 2).
      The remote completion is almost hidden from the local rank's workflow.
    \label{figure:exahype:workflow}
  }
\end{figure}

\subsection*{Parallel Implementation of ADER-DG}

%
%
Per time step, the $\mathcal{C} \circ \mathcal{R} \circ \mathcal{P}$ sequence of
cell/face/cell operations is applied to the adaptive grid which is geometrically partitioned.
We use a non-overlapping domain decomposition where the Riemann tasks along the
domain boundaries are computed redundantly by each adjacent rank.

%
%
The three ADER-DG phases translate into three types of tasks.
Prediction tasks correspond to cells, 
Riemann tasks to faces,
and correction tasks again to cells.
Out of the three task types, the predictions $\mathcal{P}$
are the computationally dominant ones.
They make up more than
97\%
of the runtime for our experiments with polynomial order $p=7$.
$p=7$ is the order we observed the best time-to-solution per accuracy for our
experiments.
While they are expensive, they work per cell,
i.e.,~have well-defined memory needs, and they are totally independent of each other.
Nevertheless, they decompose into two categories 
\cite{Charrier:19:EnclaveTasking}
of $\mathcal{P}$ tasks.
One category are tasks/cells whose faces are adjacent to a resolution
transition---that is, neighbouring cells have a different resolution---or cells
which are adjacent to a domain decomposition boundary.

The other category of $\mathcal{P}$ tasks is formed by all the remaining
$\mathcal{P}$s.
Each $\mathcal{P}$ task feeds its output into $2d$ Riemann tasks $\mathcal{R}$.
Obviously, a Riemann task $\mathcal{R}$ \replaced[id=CCPE1]{depends}{however
might depend} on more than two prediction tasks if it corresponds to a face along a resolution transition.
If an $\mathcal{R}$ task corresponds to a face along the MPI boundary, it furthermore
requires input data running through the network.
Our first category of prediction tasks---the same applies to corrections---all
have dependencies with such ``sophisticated'' Riemann tasks.
All Riemann tasks are computationally lightweight. 
After completion of all
$2d$ Riemann solves which surround one cell, the time step's final correction
task $\mathcal{C}$ is triggered.
$\mathcal{C}$ is comparably cheap, too.

%
%
As we work with a task-based formalism, our code can work with fully
non-blocking boundary data exchange.
Upon the completion of a prediction job which is adjacent to a partition
boundary, we send out the output immediately.
As our steps are phrased in tasks, we then continue with further prediction
tasks or postpone Riemann tasks which are not ready yet due to missing incoming
data.
This yields a classic MPI+X parallelisation where the boundary
exchanges do not synchronise the individual ranks
(Fig.\ref{figure:exahype:workflow}).

\deleted[id=CCPE1]{Our (baseline) code is neither bandwidth-bound all the time nor consists of
distinguishable different phases.}%
The task formalism intermixes the three compute steps
$\mathcal{C}, \mathcal{R}, \mathcal{P}$
\cite{Charrier:19:EnclaveTasking}\replaced[id=CCPE1]{.
While $\mathcal{R}$
and $\mathcal{P}$ are very cheap and thus stress the memory system, the
scheduler typically runs them parallel to some $\mathcal{C}$ tasks.
The
node's memory controllers consequently can deliver all data on time.
This%
}{and this} melange of different activities is interwoven with MPI data transfer in
the background \cite{Charrier:19:EnclaveTasking}.
If bandwidth restrictions arise, they arise as bursts towards the end of each
time step
\added[id=CCPE1]{%
when the majority of tasks has finished and all MPI communication is triggered \cite{Samfass:20:teaMPI}.
}%
The code is MPI bandwidth-demanding yet not bandwidth-bound always.

\subsection*{Optimistic time stepping with weakened temporal and spatial
constraints}

\replaced[id=CCPE1]{Generic explicit}{Explicit} time stepping for hyperbolic
equations suffers from strong synchronisation:
The outcome of one time step has to be globally reduced, as we have to
determine the admissible time step size from the CFL
condition\deleted[id=CCPE1]{, and this admissible time step size then feeds into
the ranks again}.
This is an allreduce.
\deleted[id=CCPE1]{
For the present application scenario, a dynamic time step adaption is
overengineered.
As long as the PDE remains linear and no boundary conditions such as
dynamic rupture are imposed
which dramatically change the temporal scale of the studied phenomena, the admissible time step size remains invariant.
To future-proof the code, we nevertheless realise the global CFL data exchange.
}
Once we \added[id=CCPE1]{however} assume that time step size is known or that
our code can reliably estimate the evolution of the admissible time step size a
priori\added[id=CCPE1]{---for our linear PDE with simplistic initial
conditions, this holds trivially since the admissible time step size is
invariant---}we can eliminate the strict global synchronisation of the ranks
\added[id=CCPE1]{\cite{Charrier:19:AderDG}}.
While a rank waits for an exchange of global information, incoming Riemann
data or AMR information, it can already process
$\mathcal{P}$ tasks of the subsequent time step.
Performance analysis thus has to be done carefully:
While a rank waits to complete its time step, and, hence, cannot logically
kick off the next
time step, it might still have work to do which logically belongs into this
very next time step.
\deleted[id=CCPE1]{Our methodology amplifies this effect.}

The $\mathcal{R}$ and $\mathcal{C}$ tasks have to run close to
the memory.
They are cheap\deleted[id=CCPE1]{ and notoriously memory- or
latency-bound}\added[id=CCPE1]{ and have outgoing dependencies into the
compute-heavy $\mathcal{P}$ tasks, as they couple cells with each other or
precede subsequent time steps}.
$\mathcal{P}$ tasks in contrast are candidates to be deployed to remote compute devices:
they cause the primary computational load and 
they are typically not immediately time-critical, 
at least not in the moment they are spawned. 
Other predictions are in the task
queue, and there is a high probability that further Riemann and correction steps of the previous
time step still have to be processed.
\added[id=CCPE1]{%
 Furthermore, they are compact:
 They are atomic work units whose costly computations
 require input of limited memory footprint and yield output of limited
 footprint. 
 Their arithmetic intensity is high while their input/output demands are small. %
}%
We therefore call prediction tasks
\replaced[id=CCPE1]{offloadable:}{\emph{offloadable}.}

\begin{definition}
  \added[id=CCPE1]{
   An {\bf offloadable} task is a task with high arithmetic intensity and small
   input and output data, which is furthermore not time-critical in most
   situations, i.e.~is typically accompanied by many other ready tasks.
  }
\end{definition}

%% file: 03_vision.tex

\section{Methodological vision}
\label{section:vision}

We assume that the work in our code is already reasonably distributed
via a distribution of data (grid cells, etc.) to MPI ranks.
\replaced[id=CCPE1]{%
We expect, however, that this distribution cannot lead to perfectly balanced execution times, 
because of impredictable computational load or fluctuations in system performance. 
}{%
We however accept that even 
sophisticated dynamic AMR codes will fail to achieve perfect balancing
of computing times.
This can be due to variable computational load per data entity (grid cell) in complicated models and numerical schemes but also due to the complex
hardware and performance behaviour of modern machines.
An additional motivation can be to proceed with a non-perfect load distribution, 
because of high data-redistribution costs.
}%
In MPI+X codes, 
imbalance eventually manifests
in MPI waits.
We therefore determine approximate waiting times---the measured ``wait''
is reduced by the time a rank could spend on dangling tasks that are not
critical for progress---and build a wait graph
that allows us to determine bottleneck ranks.
However, it is too late to react once ranks become idle, 
as we would essentially create further waiting times to move around tasks. 
Instead, we implement proactive task offloading in the sense that a \emph{critical rank} (identified as being too slow) will 
offload tasks to under-employed \emph{victim ranks} ahead of time (i.e., \emph{proactively}) and based on knowledge from previous time steps. 
\added[id=CCPE1]{%
 Where this offloading is too ambitious, i.e.~results do not come back on
 time, the rank reduces offloading in subsequent compute steps and ``urgently'' 
 recomputes the result itself. 
 It is reactive.%
} 
\replaced[id=CCPE1]{Our}{
Hence, our reactive} task offloading teams up with traditional data
decomposition and migration, and helps to improve load balancing.
It finally is hidden away from the code, i.e.,~it is a lightweight extension.

We further exploit that our AMR code \replaced[id=CCPE1]{runs phases
which are dominated by computations and phases where communication is
critical. Despite}{contains strongly compute-bound kernels. Hence, despite}
bandwidth access peaks\deleted[id=CCPE1]{caused by memory-bound kernels}, we
have bandwidth available in-between these peaks.
\emph{We propose to hijack 
MPI wait times and available bandwidth on ``too fast'' ranks to process
tasks that are ``stolen'' from ``too slow'' ranks.}
\deleted[id=CCPE1]{
We take compute-intense jobs from ranks that delay other ranks and
process them on ranks which wait for other ranks.
This works as we process such tasks with high
priority on the idle ranks, and thus send tasks forth and back at times when the ranks are not
suffering from bandwidth constraints.
The result is a reactive load balancing approach which is
orthogonal to, i.e.,~works on top of classic load 
balancing.
}

%% file: 04_terminology.tex

\section{Terminology}
\label{section:terminology}

Our algorithm is constructed around simple terminology and a few
definitions.
Let $N^\text{(ranks)}$ denote the number of MPI ranks.
$0 \leq i,j < N^\text{(ranks)}$ always holds for indices $i,j$.
Each rank employs $N^\text{(cores)}$ cores.
They realise the time stepping algorithm, i.e.,~process
$\mathcal{C},\mathcal{R},\mathcal{P}$. 
They notably also process all remote $\mathcal{P}$ 
tasks, i.e.,~tasks sent in by
another rank, processed locally, but then sent back.
We denote $N^\text{(tasks)}_i(t)$ to be the
number of these tasks at a certain time $t$ on rank $i$.
As tasks are migrated, spawned throughout the time step, and completed, 
$N^\text{(tasks)}_i(t)$ changes all the time.
Finally let $t^\text{(task)}$ be the time one core requires to complete
one of the offloadable tasks.
We assume they are atomic, i.e.,~run exclusively on one core at a time.
$t^\text{(task)}$ quantifies the cost of $\mathcal{P}$.
Sampling determines it \added[id=CCPE1]{introspectively: we use a moving average
to determine $t^\text{(task)}$, which implies that we assume all tasks on a given rank
to be similarly costly on average. Yet, the time window of the moving average renders
 this cost model adaptive at run time, i.e. it is reactive.}

Per time step, our code exchanges boundary data with neighbouring ranks as well
as a global time step size.
Our reactive load balancing plugs into these data exchanges.
We found it sufficient to track the global exchange only, but the concept could
be applied to the boundary exchange, too.
It thus holds also in the absence of global synchronisation.

\begin{definition}
  Our code runs into situations where a rank $i$ (logically) stops and \replaced[id=CCPE1]{cannot}{can
  not} continue until a message from rank $j$ arrives.
  Let the \textbf{waiting time} $t_{i,j}^\text{(wait)}$ be the core time
  that elapses in-between.
\end{definition}

\noindent
In a BSP-type environment (bulk synchronous processing) where a rank
forks threads, joins these again, and then finishes all data exchange,
$t_{i,j}^\text{(wait)}$ is a simple online measurement quantity:
$t_{i,j}^\text{(wait)} = N^\text{(cores)}(T_{i,j}^\text{(start)} - T_{i,j}^\text{(end)})$, where  
$T^\text{(start)}_{i,j}$ is the time stamp when rank $i$ receives the 
kick-off message of the subsequent time step from rank $j$
and $T_{i,j}^\text{(end)}$ is the time stamp when data exchange between $i$ and $j$ ends (these are
typically sends).
$t_{i,j}^\text{(wait)}$ sums up all core wait times (which are equal) and thus scales with $N^\text{(cores)}$.

In an asynchronous task environment tasks of a time step $n$ 
that are not critical to the progress of the rank 
may overlap with computations of time
step $n+1$.

We therefore reduce the wait time $t_{i,j}^\text{(wait)}$ by an
additional term:
\begin{equation}
 t_{i,j}^\text{(wait)} = \max \left( 0, N^\text{(cores)}\bigl(T_{i,j}^\text{(start)} - T_{i,j}^\text{(end)} \bigr) -
 N^\text{(tasks)}_i  t^\text{(task)}  \right).
 \label{equation:terminology:reduced-wait-time}
\end{equation}
\noindent
$N^\text{(tasks)}_i  t^\text{(task)}$ quantifies how
much of the wait time can be spent productively on handling ready tasks.
It is a crude estimate as $N^\text{(tasks)}_i$ might change dramatically throughout
this time.
Consequently, we use the $\max$ function to avoid negative wait times.

\begin{definition}
  Rank $i$ is called a 
  \textbf{critical rank} 
  if $\ \forall j: t^\text{(wait)}_{i,j}=0$ and 
  $\exists j: t^\text{(wait)}_{j,i}>0$.
\end{definition}

%
%
\noindent
A critical rank is a rank that does not wait for any other rank but 
delays at least another one.  
While there may be more than one critical rank, we usually identify the most
critical one to offload\replaced[id=CCPE1]{ tasks}{ task} from it to underloaded victim ranks:

\begin{definition}
  Let $t^\text{(wait)}_{max} = \max _{i,j} t^\text{(wait)}_{i,j}$. 
  A rank $i$ is an   \textbf{optimal victim} 
  if $\ \not \exists j: t^\text{(wait)}_{j,i}>0$ and $\ \exists
  j: t^\text{(wait)}_{i,j}=t^\text{(wait)}_\text{max}$.
\end{definition}

%
%
\noindent
An optimal victim is the rank in the system that could take up the
biggest chunk of further work without decreasing the performance, since it idles the longest.
Our goal is to make critical ranks deploy more and more tasks to optimal victims
until they cease to be critical.
For this, we introduce 
a quantity $N^\text{(offload)}_{i,j}$ per rank which clarifies
how many tasks from rank $i$ should be deployed to rank $j$.
Rank $i$ then plugs into the task spawn mechanism.
We outsource the first $N^\text{(offload)}_{i,j}$ offloadable tasks that become ready throughout a time step to rank $j$. 
In a task-based environment the ``first'' is to be read weakly, as the runtime
might reorder them.

As we work in a distributed environment with changing meshes, non-constant
numeric cost, and hardware noise, this type of non-persistent load balancing can
fail:

\begin{definition}
  An \textbf{emergency} arises for rank $j$ if $\ \exists i:
  N^\text{(offload)}_{i,j}>0$ and $t^\text{(wait)}_{i,j}>0$. 
  \label{definition:terminology:emergency}
\end{definition}

%
%
\noindent
Emergency means that a rank both deploys data to a victim rank and is
delayed by this very rank. 
This may happen when the victim rank is overloaded, if we suffer from network
congestion or if too many messages (remote tasks) stress the MPI subsystem such
that results are not sent back fast enough to the deploying rank.
As soon as we spot such an emergency, we add a rank to a black list.

\begin{definition}
  The \textbf{blacklist} is the set of ranks that may not 
  take up more work.
  We hold one blacklist per rank.
\end{definition}

\noindent
Our terminology circumscribes a greedy graph optimisation algorithm.
We establish a \emph{wait graph} over all ranks.
$t_{i,j}^\text{(wait)}$ serves as edge weight in this directed graph. 
If we mask out zero weights, the graph is sparse.
The critical rank is the
last rank along a critical path through the set of ranks.
The ``last'' edge points to the critical rank. 
Multiple critical ranks may exist.
Our goal is to remove the head from the critical path and then to continue
iteratively.

To achieve this goal, we label those ranks in the graph which are origins of
wait paths with the biggest wait time as optimal victims.
They can take up further work without slowing down the overall computation.
The determined numbers of task offloads $N_{i,j}^\text{(offload)}$ establish
a \emph{task distribution graph} on top of our rank vertices. 
It connects sinks of the wait graph with sources of critical paths.
\added[id=CCPE1]{%
 Finally, we allow ranks to compute local task outcomes even though they tried
 to offload work:
}
\begin{definition}
\added[id=CCPE1]{%
 An {\bf urgent local compute} is the computation of a task despite the fact
 that this task has been given to another rank. If we urgently recompute a task
 outcome, we neglect this task's results when they eventually drop in. 
 }
\end{definition}

%% file: 05_algorithm.tex

\section{Lightweight reactive load balancing}
\label{section:load-balancing}

%
%
Once the MPI wait times are identified, each rank $i$ maintains statistics
of $N^\text{(tasks)}_i(t)$. 
It measures all $t_{i,j}^\text{(wait)}$ and it samples
execution times to determine $t^\text{(task)}$.
Furthermore each rank has a blacklist of ranks that return remote
tasks too slowly.
All statistics are sampled over time spans through
\begin{equation}
  \tilde x = \frac{\sum _{l=0}^S \bigl( \omega^\text{(avg)} \bigr) ^l x_l}{\sum
  _{l=0}^S \bigl( \omega^\text{(avg)} \bigr) ^l }, \quad \mbox{with a
  fixed\ } 
  \replaced[id=CCPE1]{\omega^\text{(avg)} \in (0,1].}{\omega^\text{(avg)} \in
  ]0,1].}
  \label{equation:algorithm:time-series}
\end{equation}
$x_0, \dots, x_S$ are the measurements from the $S+1$ most recent time steps
($x$ being a placeholder for our quantities of interest).
We drop older measurements as further quantities alter the
moving average by less than 10\% for $\omega^\text{(avg)}
\approx 0.9$.

%
%
Our global statistics allow us to introduce various algorithms to determine
$N_{i,j}^\text{(offload)}$,
i.e., how many tasks each rank $i$ has to deploy to rank $j$.
We update $N_{i,j}^\text{(offload)}$ prior to each time step
with the most recent statistics at hand.
From hereon, newly spawned offloadable tasks on $i$ 
can be offloaded to another rank $j$
as long as they haven't exceeded our quota $N_{i,j}^\text{(offload)}$.
This definition implies that we never delegate stolen tasks
further,
i.e.~only tasks produced locally are ``stolen'' by another rank.

%
%
In order to improve parallel performance in the presence of critical ranks,
we propose different strategies.

\input{05a_reactive-lb}

\input{05b_diffusion}

\input{05b_blacklisting}

\input{05c_ccp}

\input{05d_urgent-recompute}

%% file: 05a_reactive-lb.tex
%
%
\subsection*{Reactive load balancing}

Each rank can determine its optimal
number of tasks $N_{i,j}^\text{(opt)}$ that it has to deploy
to other ranks from the global data view (Alg.~\ref{algorithm:reactive-lb}).
The iterative approach identifies the unique critical rank,
computes how much it could ``fill up'' the optimal victim,
adopts the load distribution, and then waits for the next time step's
measurements.

\input{algorithm-reactive.tex}

The algorithm's use of the term optimal in $N^\text{(opt)}$ is misleading for several
reasons:
First, it is a backward-looking optimum which derives an optimal
task distribution for the passed time step.
With AMR, the grid however might change in the present step.
Second, it relies on a weak consistency model for its input quantities,
as we use non-blocking allgather.
Some data used in the computation thus might be outdated.
Third, the quantities themselves rely on $N^\text{(tasks)}_i(t)$ which is a snapshot of
the local runtime's state.
Fourth, the formula is based upon a real-time measurement of $t^\text{(task)}$
which we determine through a weighted averaging over multiple probes.
If a core downclocks due to high energy consumptions \cite{Charrier:19:EnergyAndDeepMemory}
or failures, this does not immediately reflect in the timings.
Finally, though our formalism sticks to unique critical workers and optimal
victims, it can happen that the asynchronous balancing makes multiple ranks
consider themselves to be critical.

%% file: algorithm-reactive.tex
\begin{algorithm}
  \caption{Blueprint of reactive load balancing. 
  \label{algorithm:reactive-lb}
  }
  \begin{algorithmic}
    \Function{reactiveLB}{rank $i$}
      \State $\forall k \not=i $ exchange $t^\text{(wait)}$ (non-blocking allgather) 
      \State Compute critical rank $m$
      \If{$m=i$}
        \State Compute optimal victim $n$
        \State $N^\text{(opt)}_{i,n} \gets 0.5 t_\text{max}^\text{(wait)}/t^\text{(task)}$
      \EndIf
    \EndFunction
  \end{algorithmic}
\end{algorithm}

%% file: 05b_diffusion.tex
%
%
\subsection*{Diffusion}

There is limited sense in using $N^\text{(opt)}_{i,j}$ as it is only a guideline which
tends to rebalance aggressively.
It grabs one victim rank's MPI time completely in one rush. 
We therefore introduce per rank a relaxation factor $0.1 \leq \omega^\text{(diff)}
_i \leq 1$ and determine a task distribution from the optimal distribution
plus the current state:
\[
  N_{i,j}(k+1) = \omega^\text{(diff)} _i N^\text{(opt)}_{i,j}(k) + (1-\omega^\text{(diff)}
  _i) N_{i,j}(k).
\]

\noindent
$\omega^\text{(diff)} \approx 1$ makes the diffusion adopt the ``optimal'' task distribution
quickly, while a small $\omega^\text{(diff)}$ yields a moving
average.
The actual distribution is incrementally fitted to the optimal distribution.

We may consider our overall optimisation problem to be strongly nonconvex
and subject to fluctuations.
To reduce the risk to run into local minima with small $\omega^\text{(diff)}$, but
also to reduce the risk to introduce massive distribution fluctuations, we 
increment $\omega^\text{(diff)} \gets \min (\omega^\text{(diff)} +0.1,1)$, if 
\begin{equation}
 \forall i: \ 
 \frac{
  \sum _j | N_{i,j}^\text{(opt)}(k+1) - N_{i,j}^\text{(offload)}(k) |
 }{
  \sum _j | N_{i,j}^\text{(opt)}(k) - N_{i,j}^\text{(offload)}(k-1) |
 }
 \geq \omega^\text{(reinf)}
 \label{equation:diffusion:relaxation}
\end{equation}

\vspace{\belowdisplayskip}
\noindent Otherwise, $\omega^\text{(diff)} \gets \max ( 0.9 \omega^\text{(diff)} ,0.1) $.
\replaced[id=CCPE1]{$\omega^\text{(reinf)} \in (0,1]$}{$\omega^\text{(reinf)}
\in ]0,1]$} is fixed.

While a decrease of $\omega^\text{(diff)}$ by 10\% ensures that our diffusion
updates usually become smaller and smaller, we increase the relaxation if two subsequent
iterations drag the update with a certain intensity.
The latter typically happens if ranks enter the blacklist:

%% file: 05b_blacklisting.tex
%
%
\subsection*{Blacklisting}

Our load balancing strategies can run into situations 
where they overbook ranks and thus slow down
the overall computation -- despite the damping of the updates through $\omega^\text{(diff)}$.
The paragraph following Definition \ref{definition:terminology:emergency}
enlists reasons for this and introduces
blacklists that accommodate this problem.

Whenever a victim rank does not deliver the result of a
stolen task back fast enough, the origin rank identifies this emergency and adds the
victim rank to its local blacklist. 
Blacklists are subject to our non-blocking all-gather communication and thus shared globally.
The update of the local load distribution sets $N^\text{(opt)}=0$ for any blacklisted
communication partner.
In a diffusive world, this triggers a gradual retreat from overbooked ranks.

We found it valuable to use an annotated blacklist set where each entry holds
a weight.
As long as emergencies arise for a particular rank, its blacklist value is
incremented by one.
After each rebalancing round, we decrement the weight by 10\%.
Blacklist entries with a weight below $0.5$ are eventually removed from the
blacklist.
We avoid oscillations:
If a rank has entered the blacklist, it remains on this list for a while to avoid
that it is immediately rebooked after enough tasks have been retreated.

%% file: 05c_ccp.tex
\subsection*{(Reduced) Chains-on-chains partitioning}

Diffusion yields a slow process.
This is especially true at startup if we start from an ill-suited domain
decomposition.
Furthermore, no iterative technique is safe from running into 
local minima.
It is hence reasonable to benchmark against an ``optimal'' task distribution
that is computed for a given grid setup.
The term optimal however is to be chosen carefully, as any precomputation relies
on an a-priori cost model which can only approximate the actual machine
behaviour.
We use a uniform cost model for $\mathcal{P}$ which neglects data transfer cost.

Chains-on-chains (CCP) partitioning~\cite{Pinar:04:CCP} is one approach to
determine good task distributions.
It can be defined as partitioning of
a 1D chain of $\sum _i N^\text{(tasks)}_i$ tasks into $N^\text{(ranks)}$
partitions such that the bottleneck load (maximum load assigned to a rank) is minimized. 
With uniform cost per task, the CCP problem reduces to a much simpler problem: 
we only need to ``cut the chain'' of tasks into $N^\text{(ranks)}$ equally sized pieces, i.e., the bottleneck load will then be equal to the average load over all ranks ($\pm$1 task).
The number of tasks per rank is known after the initial mesh was built on every rank. We use a single collective allgather step to distribute
this information among all MPI ranks. 
Every rank then solves the reduced
CCP problem using a simple search algorithm. This results in a new unique partitioning that defines
how many tasks every rank needs to give to other ranks such that the new load on every rank is rendered equal to the average load.

%% file: 05d_urgent-recompute.tex
\subsection*{Urgent local compute}

\added[id=CCPE1]{
 We offload solely ready tasks. 
 If results of offloaded tasks come back too late, blacklisting becomes active.
 This is a reactive strategy to accommodate unexpected performance breakdowns.
 However, it remains a proactive mitigation and does not moderate the
 immediate performance penalty arising from a lack of task results.
 If a rank experiences a performance drop, blacklisting and task
 reassignment react to this change of performance two or three time steps later.
}

\added[id=CCPE1]{
 We therefore propose an extension of our scheme that tackles sudden performance
 drops:
 Whenever our algorithm offloads a task to another rank, a local copy of this very task is
 stored and kept on the origin rank as well.
 We call this task a \emph{local recompute task}.
 If we run into an emergency, we continue to blacklist.
 Instead of an idling wait, we however compute the outcome of the task we are
 waiting for locally.
 We handle the local recompute task.
 We eventually can proceed even though task results still have not come
 back.
 The underlying offloaded task is internally marked as recomputed.
 When its result comes back, we throw it away, as we have already determined the
 task outcome locally.
}

\added[id=CCPE1]{
 Urgent local recomputes are accompanied by some local overhead, as we have to
 realise some additional bookkeeping.
 Its most important implication is that it changes the blacklisting behaviour:
 Whenever a rank waits for offloaded task results from ranks $i$ and $j$, a
 realisation without urgent recomputes blacklists first $i$, waits for the
 result of $i$ and then checks $j$.
 This gives $j$ more time to get its results back.
 With urgent recomputes, there is a higher probability that both $i$ and $j$ are
 blacklisted.
 We found it thus advantageous to explicitly mask out such situations, i.e.~to
 stop any blacklisting after one emergency until the underlying emergency's
 rank has finally got its results back.
}

%% file: 07_implementation.tex
\section{Implementation}
\label{section:implementation}

The success of our reactive, lightweight load balancing
hinges upon an efficient, low-overhead realisation. In particular,
we rely on fast task migration for an irregular, a-priori unkown
dynamic communication pattern. We found that prioritized task processing and
full overlap of task communication are essential.
The latter requires dedicated attention from a technical standpoint,
as sufficient ``progression'' of MPI messages needs to be ensured.

Our implementation is based on 
Intel's Threading Building Blocks (TBB)
\cite{Reinders:07:TBB}
which we extended by a custom priority mechanism:
We employ as many real TBB tasks as we have cores per rank. 
These TBB tasks process (consume) our own, logical tasks managed through TBB's priority
queue.
We found this solution to outperform the native TBB priorities.

\input{07a_task-lifecycle}
\input{07b_probing}

\input{07b_progression}
\input{07c_measurements}

%% file: 07a_task-lifecycle.tex

\subsection*{Task lifecycle and decision making}
Our runtime distinguishes three types of tasks: High priority tasks, low
priority tasks and offloadable tasks. 
Low priority is the default.
The offloading hooks into the actual creation of offloadable tasks.

\input{algorithm-spawn}

The hook makes the decision whether a task is enqueued locally or can be offloaded.
For this, it combines three criteria (Alg.~\ref{algorithm:spawn}):
The task has to be offloadable\deleted[id=CCPE1]{ (have reasonable high
arithmetic intensity and cost)}, there have to be more than $C$ tasks in the
local task queue, and there has to be a victim rank.
We store an atomic counter for each target rank $j$ of $i$.
It counts how many tasks still might be offloaded to rank $j$. 
It is updated per time step by the load balancing. 
If a task is given away, the respective counter is decremented.
As we may need to offload tasks to multiple victim ranks, victims are 
selected in a round-robin fashion. 
Round-robin ensures that victim ranks can start to process offloaded
tasks as soon as possible.

We exploit that each task is ready when it
is spawned. 
For codes with task dependencies, the offload decision would need to hook into the
transition of a task into ready.
Giving away tasks too aggressively can lead to starvation of rank-local task consumers. 
We face a classical consumer-producer challenge: 
The code spawns tasks only at a certain speed and puts them into the job queue.
Besides the limited speed, not all tasks are offloadable.
Hence, if we give away tasks too aggressively to other ranks---which act as 
additional consumers---the task job queue may run out of tasks for local
processing. 
To avoid this, we only offload a task if enough tasks remain available
to keep the local task consumers busy. 
This guarantees optimal utilization of both local and remote resources.

Tasks that are offloaded logically split up into two tasks:
While the actual task is sent away and computed remotely, we logically insert a single
receive task for all offloaded tasks. 
The runtime will poll this receive task as part of the standard task
processing.
Once a remote task starts to send its results back, the receive task finalises
the corresponding MPI receives and cleans up all data structures.
The task offloading itself is not visible to the application.

To ensure that offloaded tasks are sent back early, i.e.,~to ensure that we make
optimal use of the network, we issue offloaded tasks with high priority. 
They are thus computed prior to local tasks. 

%% file: algorithm-spawn.tex
\begin{algorithm}
  \caption{Spawn process of a ready task on rank $i$. 
  At the start of each time step $\hat N^\text{(offload)}_{i,j}
  \gets N^\text{(offload)}_{i,j}$.
  \label{algorithm:spawn}
  }
  \begin{algorithmic}
    \Function{spawnTask}{rank $i$, task $x$}
      \State $notStarved = N^\text{(tasks)}>C$
        \Comment Avoid rank starvation 
      \If{$notStarved \wedge canOffload(x)$}
        \State $j = pick arg_k \{\hat N_{i,k} > 0\}$
          \Comment Round robin
        \If{$j \not= \bot $}
          \State $\hat N^\text{(offload)}_{i,j} \gets \hat N^\text{(offload)}_{i,j}-1$
            \Comment Atomic
          \State Send $x$ to rank $j$
        \EndIf
      \Else
        \State Enqueue $x$ with low priority
      \EndIf
    \EndFunction
  \end{algorithmic}
\end{algorithm}

%% file: 07b_probing.tex
\subsection*{Creating the reactive communication graph}

Predictive load balancing algorithms, such as CCP, use a dedicated
synchronization step where load balancing meta-information is exchanged.
As a result, all communication partners are known prior to the actual
computation and \texttt{MPI\_Irecv}s
can be posted at the time of the load migration. 
In our reactive scheme, 
we do not explicitly exchange
meta-information, 
the communication pattern changes frequently 
and the round robin task distribution
makes it impossible to predict the exact data flow as well as the
number of messages to be transferred.
Finally, tasks are to be sent out as soon as possible, i.e.,~we may not
aggregate tasks.

Our algorithm resembles a one-sided data exchange model where many small tasks
are ``put'' to another rank and have to ``trickle through'' while the numerical
algorithm is running.
Without a mutual a-priori agreement on the communication pattern and the size
of receive windows, i.e.,~the data cardinality, we however issue one asynchronous
data send per task that is to be offloaded, and we use \texttt{MPI\_Iprobe} to
detect tasks that are to be received.

This yields many small non-blocking data transfers.
Their efficient realisation, i.e.,~the quick establishment of data flows---we may
assume that they are large enough to prohibit eager buffering---is very important as we have to release critical
ranks from work.
We make an additional task realise the \texttt{MPI\_Iprobe} pick ups.
It polls MPI, establishes incoming data connections, i.e.,~launches receives, and
eventually reschedules itself after all the other ready tasks.
Therefore, the task's mean time between activation automatically depends on the
load:
The longer the task queue the longer it will take until the probing is executed
again.
In phases of high computational load, CPU time is mostly dedicated to computation. 
In phases of low computational load, in particular whenever a rank is
underloaded and thus a potential victim, rescheduling ensure that
tasks are received quickly.
We busy-poll MPI.
Once a remote task completes, its host rank, i.e.,~the victim,
triggers a send back.
It is another non-blocking send which is eventually
picked up by the probes on the task's origin.

%% file: 07b_progression.tex
\subsection*{MPI progression}

Our code stores all pending sends and receives, i.e.,~the \texttt{MPI\_Request}
handles, in a central broker (``request manager'').
They are held FIFO. 
A central difficulty with many non-blocking MPI messages and a dynamic exchange
pattern, however, is progressing messages in the background. 
Issuing solely \texttt{MPI\_Isend} does not ensure that the actual message
transfer occurs fully in the background without any further CPU involvement
\cite{Hoefler:08:SacrificeThread}.
We cannot be sure that MPI makes sufficient progress.

One possible remedy is to sacrifice a thread for asynchronous MPI progression. 
However, neither do all MPI implementations support dedicated progression
threads, nor did we succeed to use them robustly on our test system, nor are we
eager to sacrifice a whole thread.
Even if it is pinned to a hyperthread, a progression thread tends to pollute the
runtime characteristics and caches.

We therefore implemented a \texttt{progress} task (similar to
\cite{Buettner:2013}) which uses 
\texttt{MPI\_Testsome} on the request manager's request queue to make progress
on outstanding MPI requests.
In line with the polling, the task is started prior to the first time step and
reschedules itself.
Its rescheduling policy is different to the polling:

Requeuing at the end of the ready queue turns out to be insufficient when a
critical rank sends away tasks aggressively to victim ranks. 
A critical rank is per definition overloaded, i.e.,~has a long task queue.
Too little investments into MPI progress yield late receives on the victim side. 
The progression task therefore forks an additional very high priority task
if there are outstanding send requests. This task is terminated once
no more outstanding send requests are remaining. On the receiving side,
i.e., on an optimal victim rank, a very high priority progression task is spawned 
if there are outstanding receive requests.
The latter is terminated once the receiver's set of active senders is empty.

\added[id=CCPE1]{
 Packing all tasks outsourced to one victim rank into one big message
 \cite{Jofre:15:ParallelLB} could mitigate the need for aggressive,
 manual MPI progression, since fewer (larger) messages are exchanged.
 It can however delay the outsourcing on the sender side:
 If non-migratable tasks ``suddenly'' are inserted into the local task graph,
 the assembly of a particular set of outsourced tasks can be significantly
 delayed. 
 Such situations arise, if a mesh traversal has to realise dynamic adaptive mesh
 refinement early throughout a time step within its local domain.
 On the receiver/victim side, a collection of the task outcomes that are to be
 returned can imply that the outsourcing rank recomputes outsourced data locally
 even though it would have been available on time.
 We reduce the communication to computation overlap.
}

%% file: 07c_measurements.tex
\subsection*{Data calibration}

Our reactive load balancing algorithm relies on online
performance measurements which is distributed using non-blocking collective 
communication (\texttt{MPI\_Iallgather}).
With real time stamps, it is clear that an effective zero wait time does not
manifest in a zero time span. 
We thus determine a threshold 
$t_\text{min} = 0.95 \cdot \min_{i,j}t_{i,j}^\text{(wait)} + 0.05 \cdot
\max_{i,j}t_{i,j}^\text{(wait)}$ for each rank, and drop all times below $t_\text{min}$.

Nevertheless, some data remain biased:
The wait time as defined in \eqref{equation:terminology:reduced-wait-time} notably suffers
from snapshotting effects in MPI.
Before we bookmark $N^\text{(tasks)}_i(t)$, we run an additional instance of our polling
task. 
It otherwise might happen that \eqref{equation:terminology:reduced-wait-time} assumes that no tasks were there
even though they roam in MPI.
This would eventually yield wrong timings and input into our algorithm.

We finally point out that (\ref{equation:terminology:reduced-wait-time}) is a
very idealised machine model: 
Our formula does anticipate that pending tasks can be done while we wait for
incoming MPI messages and cores thus do not idle, but the formula does not
distinguish where these remaining ready tasks come from.
If no tasks are stolen, it is reasonable to assume a fixed cost $t^\text{(task)}$ for
a homogeneous set of pending tasks. 
If some of these tasks however are stolen tasks, their cost is higher, as we
eventually have to send these tasks back.
The vanilla version of (\ref{equation:terminology:reduced-wait-time})
underestimates the local load, thus yields too high wait times, and eventually
traps the reactive load balancing in an overbooking of victim ranks.
It is therefore reasonable to reduce the local load further by a penalty which
correlates linearly to the number of received tasks (which is encoded in our
request manager).

%% file: 08_results.tex

\section{Results}
\label{section:results}

We benchmark our code on SuperMUC phase 2 and Super\-MUC-NG
at the Leibniz Supercomputing Centre
(LRZ).
Each of phase 2's two-socket nodes contains two 14-core Intel
Xeon E5-2687 v3 (Haswell) CPUs.
Throughout the experiments, they have been clocked at 
2.3~GHz.
Infiniband FDR14 connects the individual nodes with a non-blocking pruned 4:1
tree.
SuperMUC-NG hosts $2 \times 24$ cores of the Intel Xeon 8174 (Skylake)
 generation \added[id=CCPE1]{per node}, which are clocked at 2.3~GHz and are connected through Intel Omni-Path.
All shared memory parallelization relies on Intel's Threading Building
Blocks (TBB)~\cite{Reinders:07:TBB} while Intel's C++ compiler translated all codes.
We use the 2018 generation of both tools on SuperMUC phase 2 and the 2019 generation of both tools on SuperMUC-NG.

\begin{figure}
\centering
\includegraphics[width=0.8\textwidth]{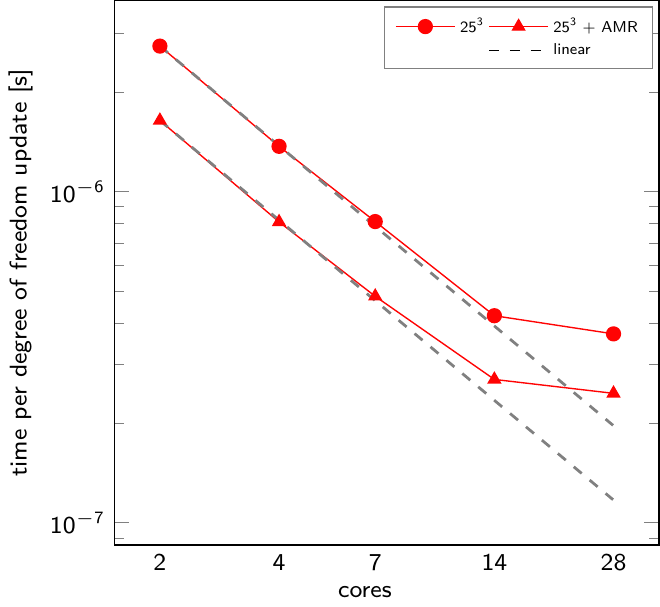}
\caption{
  \added[id=CCPE1]{
  Shared memory parallel efficiency of our baseline code on one node without any
  offloading. 
  We start from a $25 \times 25 \times 25$ grid ($15,626$ cells) and then add
  one additional level of (static) AMR ($58,525$ cells).
  }
}
  \label{figure:shared_mem_efficiency}
\end{figure}
 
Benchmarking with the baseline code \replaced[id=CCPE1]{reveals}{has revealed}
that we achieve a \added[id=CCPE1]{high} shared memory efficiency%
\deleted[id=CCPE1]{well above 93\%} on one socket
\replaced[id=CCPE1]{(Figure~\ref{figure:shared_mem_efficiency})}{Table~\ref{tab:shared_mem_efficiency}}
for a regular grid.%
\deleted[id=CCPE1]{, while the four core setup even yields an efficiency of
100\%.} Performance deteriorates once we exceed 14 cores
\replaced[id=CCPE1]{as NUMA effects kick
in\cite{Charrier:19:EnergyAndDeepMemory}.} {(a more detailed study on single-node performance, with emphasis on memory 
performance and scalability restrictions, was presented in previous work
\cite{Charrier:19:EnergyAndDeepMemory}).}

We therefore typically run multiple-of-two ranks per node.
For adaptive grids, our \replaced[id=CCPE1]{scalability}{efficiency} is%
\deleted[id=CCPE1]{only} slightly worse.
\replaced[id=CCPE1]{%
 Our task parallelisation exposes some freedom to move
 tasks around.
 With AMR, the task cost are more heterogeneous as interpolation and restriction
 tasks enter the system, too.
 This causes the slightly inferior scalability on one socket and an
 amplification of the NUMA effects.
 AMR's better cost per degree of freedom here is classic weak scaling effect.
 AMR management overhead is amortised by the higher degree of freedom count.
 The present setup uses a static adaptive mesh, i.e.~a mesh where we use a
 regular grid and then add one more level to some mesh cells.
 The results for dynamically adaptive grids do not differ qualitatively
 \cite{Charrier:19:EnclaveTasking}.
}{
 On one socket, our task parallelisation exposes some freedom to move
 tasks around.
}

\input{08a_illustration}

\input{08b_comparison}
\input{08c_acceleration}
\input{08f_emergencies}

\input{08d_speedup}

%% file: 08a_illustration.tex
\subsection*{Task and wait graph characterisation}

%
%
We kick off our work distribution experiments with a showcase to
illustrate the algorithms' behaviour\added[id=CCPE1]{ for stationary grids. No
urgent recomputes are employed so far}.
The setup uses a $25\times 25\times  25$ grid 
(leading to a problem size of 72 Mio degrees of freedom) 
on a single Haswell node hosting eight MPI ranks.
The load decomposition with eight ranks has to be
imbalanced.
We make the code dump all task outsourcing
and wait time information and use these data to 
extract the graphs underlying our algorithmic mindset.

\begin{figure}[hp]
  \begin{subfigure}[b]{0.5\linewidth}
    \includegraphics[width=\linewidth]{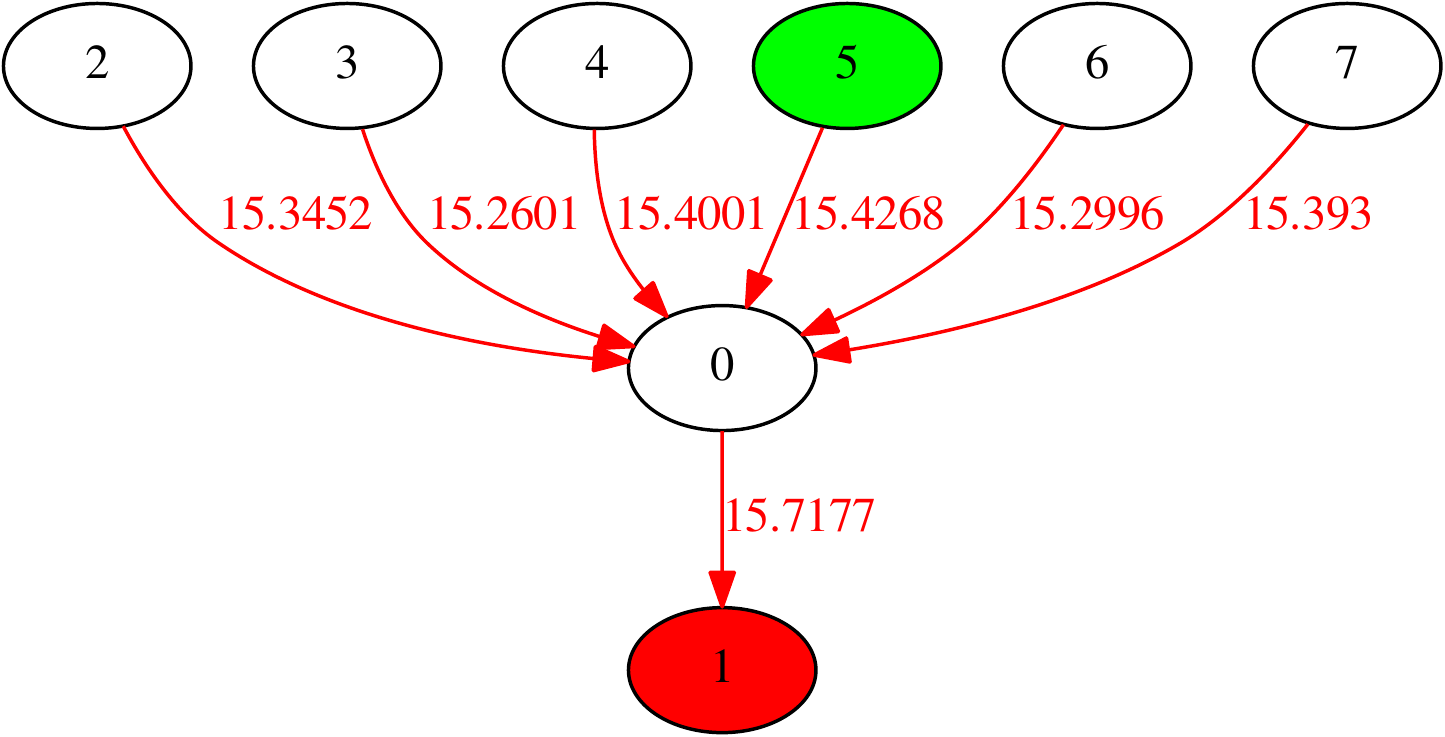}
    \subcaption{Diffusion after 7 time steps.} 
    \label{figure:results:illu7}
  \end{subfigure}%
  \begin{subfigure}[b]{0.5\linewidth}
    \includegraphics[width=\linewidth]{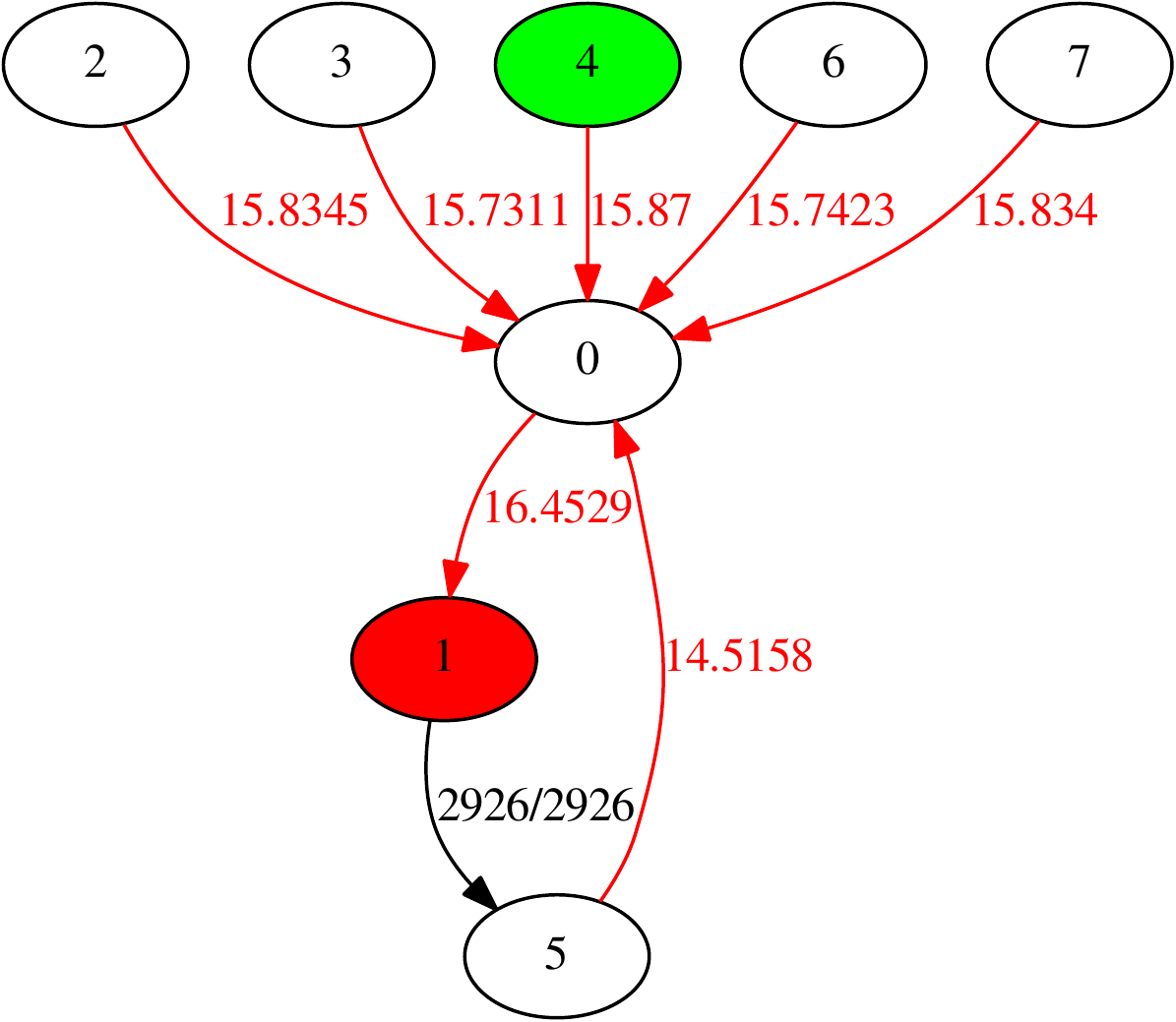}
    \subcaption{Diffusion after 10 time steps.} 
    \label{figure:results:illu10}
  \end{subfigure}
  \begin{subfigure}[c]{\linewidth}
    \begin{center}
    \includegraphics[width=0.65\linewidth]{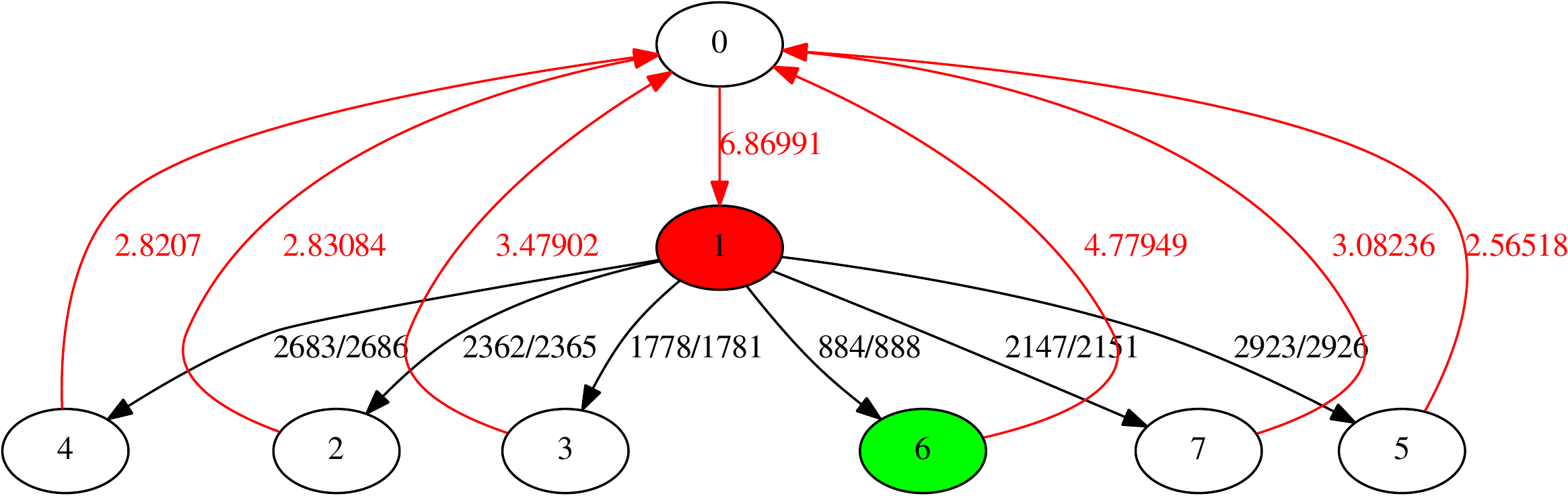}
    \subcaption{Diffusion after 200 time steps.} 
    \label{figure:results:illu200}
    \end{center}
  \end{subfigure} \\
  \begin{subfigure}[c]{\linewidth}
    \begin{center}
    \includegraphics[width=0.65\linewidth]{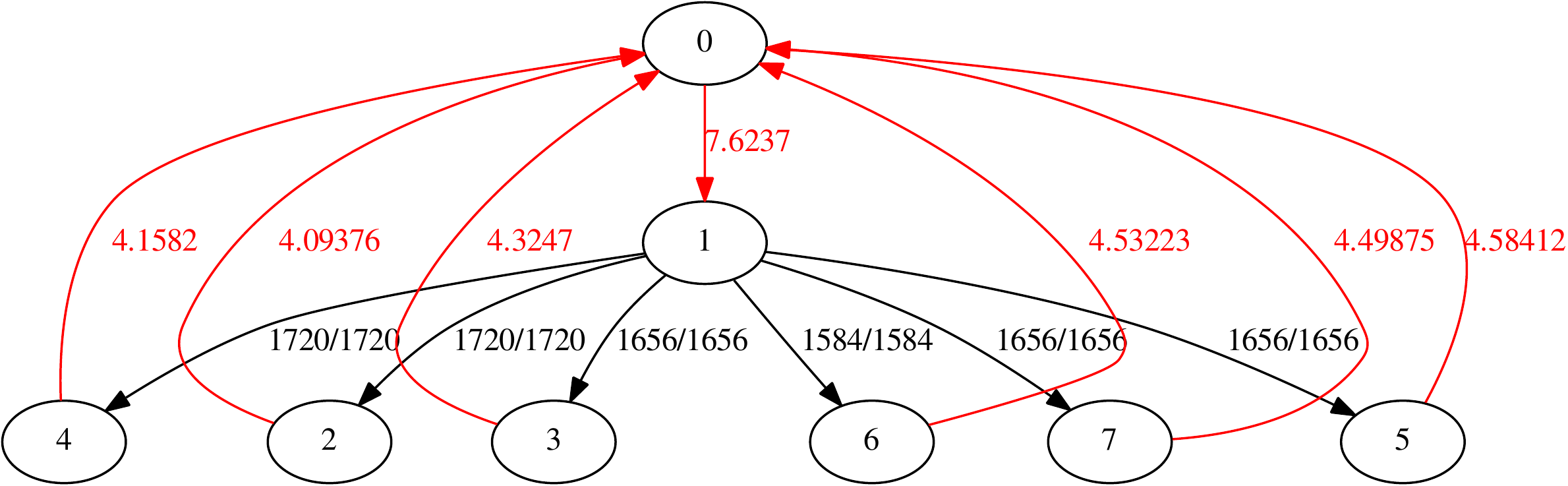}
    \subcaption{CCP after 200 time steps.} 
    \label{figure:results:illuccp200}
    \end{center}
  \end{subfigure}
  \caption{
    Wait and task distribution for the diffusive algorithm (~(a)-(c) ) 
    and our task offloading using only the\deleted[id=CCPE1]{static} CCP guess (~(d) )
    for eight ranks. 
    The critical rank is highlighted in red, the optimal victim in green.
    Red edges are wait times in seconds, black edges
    illustrate task offloading.
    The two given task numbers denote offloaded tasks vs.~maximum tasks a rank would have been
    allowed to offload.
    \label{figure:results:illustration}
  }
\end{figure}


The graphs (Fig.~\ref{figure:results:illustration}) reveal
that there is an overbooked rank 1 which delays our main time stepping loop running on rank 0. 
As rank 0 has to wait for rank 1, it in turn throttles the remaining six ranks
that wait for a kick-off of the next time step.
Such knock-on effects explain that our wait graphs will always resemble
tree or forest graphs.
It is reasonable to address the tails of the wait graphs to bring down the
runtime iteratively.

The diffusive scheme, here ran with fixed $\omega ^\text{(diff)}=1$, starts to
gradually outsource tasks from the overbooked rank to all other ranks.
The optimal victim role is passed on from one rank to the other (compare
Fig.~\ref{figure:results:illu7}
and~\ref{figure:results:illu10}) until all possible victims have been selected.
The load distribution then stabilizes and is subsequently only altered by a small number of tasks. 
CCP yields a very similar task distribution scheme for the present setup.
Our reactive diffusion thus is consistent in a numerical sense.
Overall, CCP seems to balance more evenly across the ranks 2--7, while the
number of offloaded tasks per rank is lower.

The task graphs' black labels lead to a further interesting observation.
Both balancing schemes derive a maximum number of tasks $N^\text{(opt)}$ per rank
which determines how many of these tasks can be given away.
As tasks however first have to be created---an effect that amplifies
for AMR where the task graph is unknown prior to the time step and dynamic
refinement and coarsening can delay the creation of some tasks as the grid
first has to be adopted---not all ranks fully exploit their task quota.

\begin{figure}
\centering
  \centering
    \includegraphics[width=0.7\linewidth]{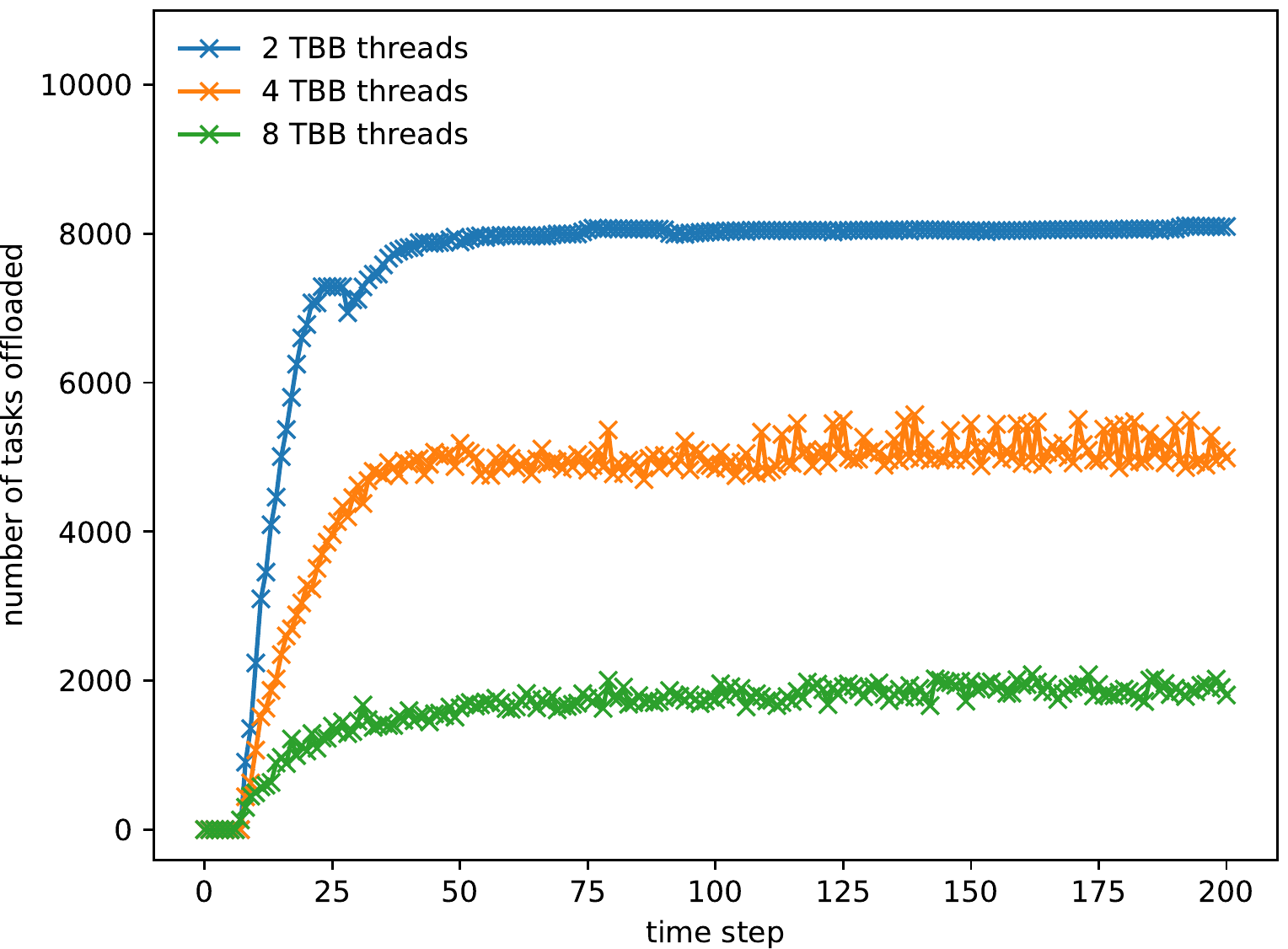}
  \caption{
    Run with 16 ranks (one MPI rank per node) where we vary the number of threads available to each
    rank.
    \label{figure:results:tasks-per-rank}
  }
\end{figure}%

\begin{figure}
\centering
    \includegraphics[width=0.7\linewidth]{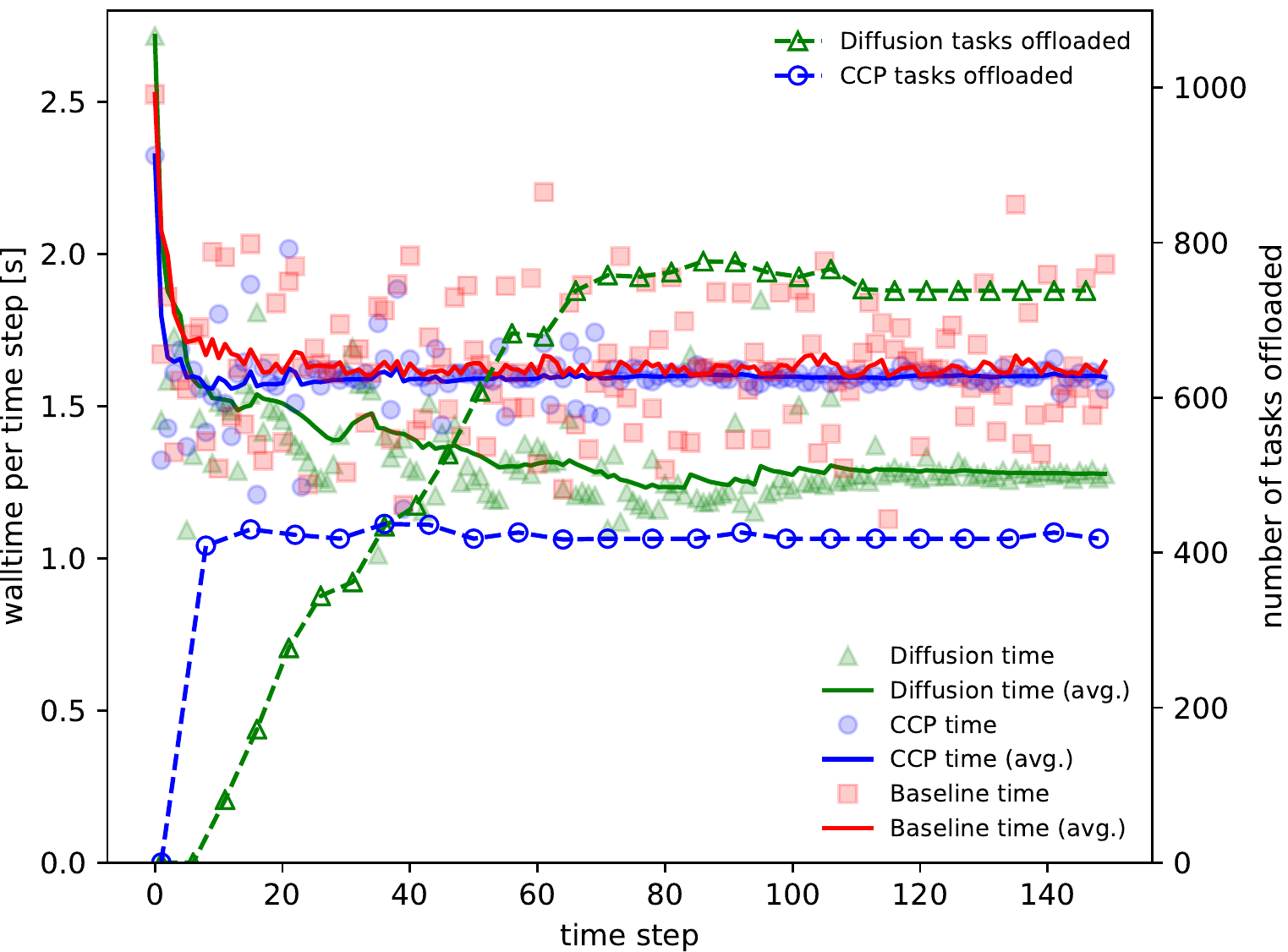}
  \caption{
    Comparison of CCP and diffusion with $\omega^\text{(diff)}=0.5$ to the
    baseline runtime.
    Per test, we present both the number of tasks that are offloaded and
    the runtime as gliding average as
    phrased by (\ref{equation:algorithm:time-series})
    \added{(SuperMUC phase 2)}.
    \label{figure:results:comparison-baseline-algorithms}
  }
\end{figure}

We continue to investigate this effect in further experiments where we use 16 MPI ranks distributed to 
16 nodes and parametrise the number of cores available to each rank.
We see ranks
deploying the fewer tasks the more cores they have locally available
(Fig.~\ref{figure:results:tasks-per-rank}).
For many codes deployed to multi-socket systems,
it is reasonable to use more than one rank
per node.
This reduces NUMA effects%
 \cite{Charrier:19:EnergyAndDeepMemory}
.
Our reactive load balancing supports such a strategy.
Otherwise, too many local cores have to be kept busy.
These two technical advocates for multiple ranks per node finally are supported
by the observation that more ranks give the domain decomposition more
degrees of freedom how to distribute the mesh.

%% file: 08b_comparison.tex
\subsection*{Comparison of baseline algorithms for an almost balanced mesh}

We continue with a comparison of our two lightweight redistribution algorithms,
CCP and reactive diffusion,
to the baseline code performance.
Again, the grid is fixed to $25 \times 25 \times 25$.
We employ $28$ ranks in total.
Due to the dominance of the $\mathcal{P}$ tasks, it is reasonable to assess
the balancing quality in terms of the distribution of the space-time predictors:
The lightest eight ranks host 512 of these tasks, while the heaviest rank hosts
729 $\mathcal{P}$s.

The measurements reveal (Fig.~\ref{figure:results:comparison-baseline-algorithms}) how hard it is to balance
and tune our baseline code---a property we consider to be prototypical
for modern, task-based simulation codes:
The runtimes per time step scatter significantly
even though this is a regular grid setup without AMR.
Closer inspection uncovers that the runtime does not randomly fluctuate but
exhibits an oscillation-type pattern.
Our code yields a task graph where individual tasks are optimistic. 
It is thus possible to bring tasks forward and to compute them in the
(logically) previous iteration already.
This leads to oscillating behaviour: 
One iterate finishes quickly. 
Tasks of the follow-up time step are set ready but not processed before the
iteration reports ``done'' to the other ranks and completes its boundary data
exchange.
The subsequent iterate now has to process all of its tasks.
At the same time, its task processing already spawns tasks of the subsequent
iteration.
Some of them are processed straight away as they sit in the ready queue.
Compared to the previous time step, the present time step thus lasts longer.
The fact that it already computes (some of) the tasks of the next iteration in
turn makes the subsequent iterate finish fast again.
We end up with oscillations.

Both of our balancing techniques reduce the oscillations.
While is reduces the noise/scattering,
CCP yields a time-averaged time per step which is hardly better than the
runtime of the baseline code.
CCP's quasi-static ``re''-balancing or ``on-top''-balancing fails
to improve the performance.
The reason is that CCP is agnostic of the real time
behaviour of the multithreaded code.
A positive insight is that the offloading's
overhead is small, as we do not loose performance with CCP. 
CCP is not
slower than the baseline.

The diffusive approach clearly outperforms CCP. 
It reduces the runtime almost monotonically and, once
converged, brings the runtime per time step down from approximately 1.6s per
step to around 1.2s.
The measurements support our decision to use work
with (\ref{figure:results:comparison-baseline-algorithms}) for measurements, and
we observe that the diffusion, anticipating real hardware behaviour, calls
for convergence acceleration techniques.
The improvement of the runtime is slow.
Most importantly, the diffusion is faster than the static-cost model of CCP.
Taking real measurements into account is important.

%% file: 08c_acceleration.tex
\subsection*{Convergence acceleration}

\begin{figure}
\centering
    \includegraphics[width=0.7\linewidth]{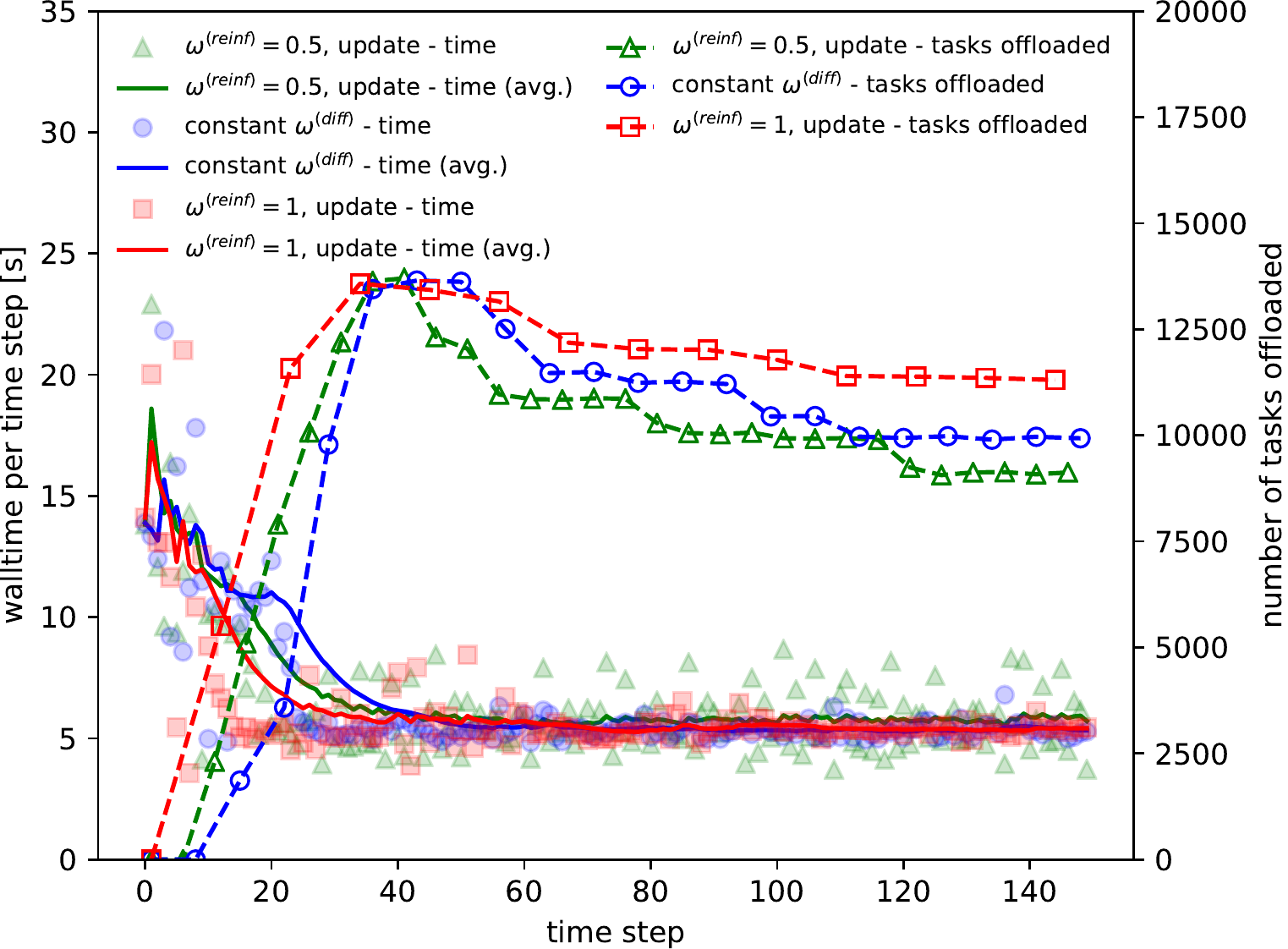}
  \caption{
    Runtime comparison of three executions of the diffusive algorithm. 
    All executions start with $\omega ^\text{(diff)}=1$. 
    Two of them alter this diffusion parameter according to
    (\ref{equation:diffusion:relaxation}).
    We use a regular grid with $25 \times 25 \times 25$ cells on a single node
    hosting $14$ ranks
    (SuperMUC phase 2).
    \label{figure:acceleration:undamped-start}
  }
\end{figure}

\begin{figure}
  \centering 
  \includegraphics[width=0.7\linewidth]{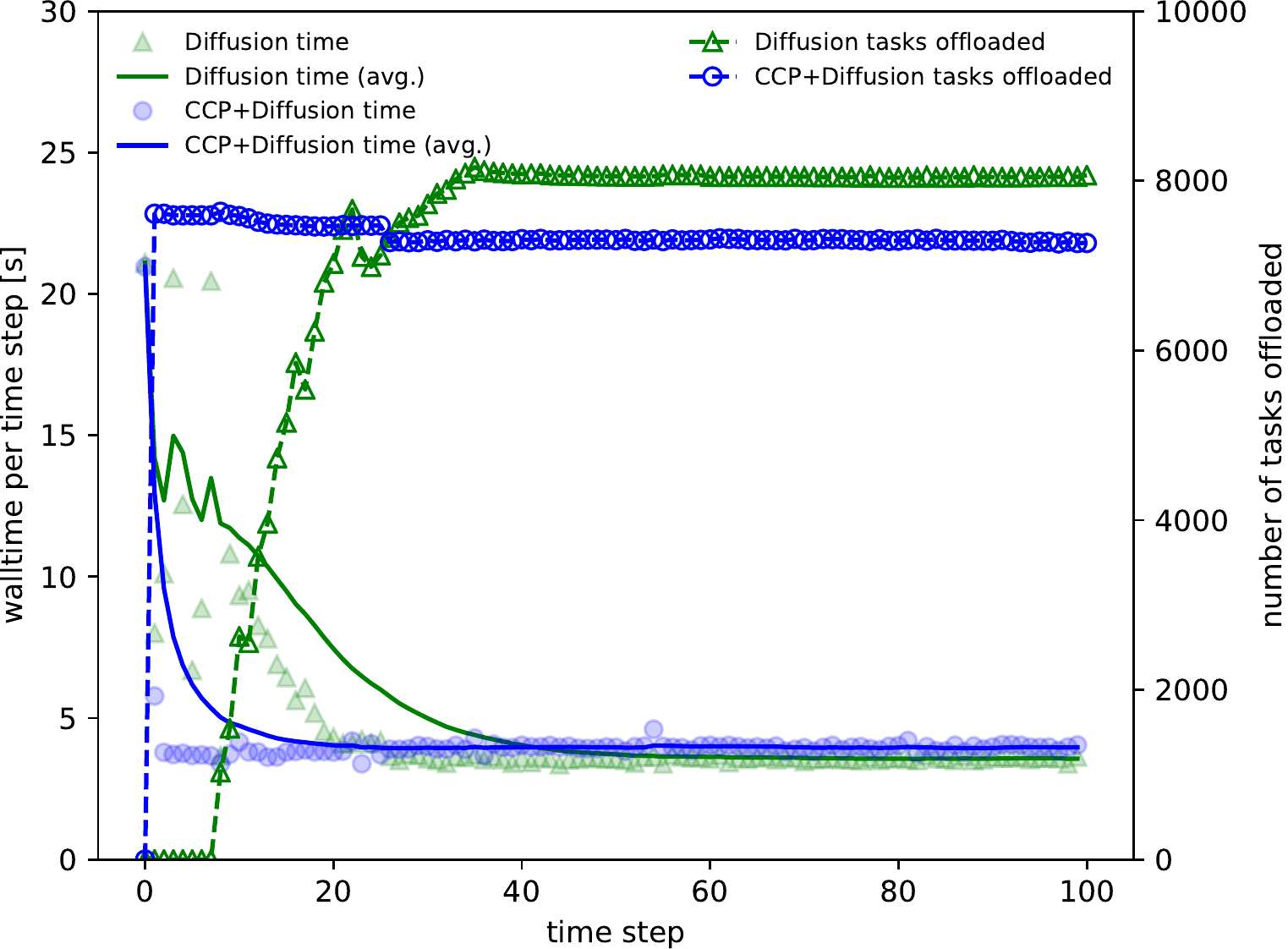}
  \caption{
    Runtime comparison of the diffusive algorithm with and without using CCP as an initial guess. 
    We always initialise $\omega ^\text{(diff)}=0.5$
    (regular grid with $25 \times 25 \times 25$ cells on a single node of SuperMUC phase 2
    hosting $14$ ranks).
    \label{figure:acceleration:ccp}
  }
\end{figure}

Our work proposes to accelerate the damping update in
(\ref{equation:diffusion:relaxation}) through a reinforcement technique. 
For all diffusion-based approaches, i.e.~for any choice of 
$\omega ^\text{(diff)}$ and with and without an adaption of this value according to
(\ref{equation:diffusion:relaxation}), the runtimes per time step eventually
converge towards a similar value.
Measurements in Fig.~\ref{figure:acceleration:undamped-start} show that
our reactive approach tends to ``over-balance'',
unless we reduce $\omega^\text{(diff)}$ in each time step.
Over-balancing manifests in a large number of over offloaded tasks
which trigger an emergency and thus induce a steep decline of tasks afterwards.
It is only $\omega^\text{(reinf)}=1$, where no emergency is triggered and we thus do
not observe a rapid decrease of offloaded tasks.
With $\omega^\text{(reinf)}=1$, both overshooting and retreat are damped, as
(\ref{equation:diffusion:relaxation}) triggers an almost monotonous decay of
$\omega ^\text{(diff)}$.
For $\omega^\text{(reinf)}=0.5$, the diffusion parameter is not immediately
decreased.
It even increases over the first few steps.
And once a rank hits the blacklist, (\ref{equation:diffusion:relaxation})
increases $\omega^\text{(diff)}$ of the rank which caused the blacklisting again.
The rank consequently retreats quickly.

Both choices of a dynamic change of $\omega ^\text{(diff)}$ outperform a static
diffusion constant.
By means of a rapid reduction of runtime, a quick reduction of $\omega
^\text{(diff)}$ is the best choice after a massive rebalancing step which is induced
here by the initial domain decomposition but also might result from dynamic AMR.
We however do observe that quick re-increases of $\omega ^\text{(diff)}$ due to small
$\omega ^\text{(reinf)}$ might be reasonable if a small number of redistributed tasks
is an objective, too.
The reinforcement acts as additional penalty to the underlying
optimisation problem which takes task offloading cost into account.

If we repeat our benchmark with 14 ranks ($\omega ^\text{(diff)}=0.5$ and
$\omega ^\text{(reinf)}=1$), and benchmark our reactive scheme 
against CCP, 
we see CCP yield an aggressive initial task decomposition (Fig.~\ref{figure:acceleration:ccp}).
This is qualitatively in line with Fig.~\ref{figure:results:comparison-baseline-algorithms}:
CCP's time per timestep is 
reduced much faster compared to the diffusion-only run. 
Reactive diffusion however is superior to CCP in the end as it takes the real
behaviour of the machine into account.
We emphasize that these experiments use CCP to determine the initial
distribution but then let diffusion take over.
CCP speeds up the initial distribution, but it also seems to steer the
reactive approach into a local minimum, and diffusion fails then to improve upon
this load balancing further.

%% file: 08f_emergencies.tex
\subsection*{Sudden performance drops}

\added[id=CCPE1]{
 All experiments so far employ stationary grids. 
 As a result, the task distribution converges towards a steady state.
 However, one might argue that a proper (re-)balancing of the workload
 would be more effective in this case.
 Yet, this holds if and only if the typical rebalancing cost including all data
 movements is significantly lower than 10--20 time steps, as this is the
 characteristic timescale of our reactive load balancing to yield good time to
 solution ratios.
}

\begin{figure}
\centering
\includegraphics[width=0.7\textwidth]{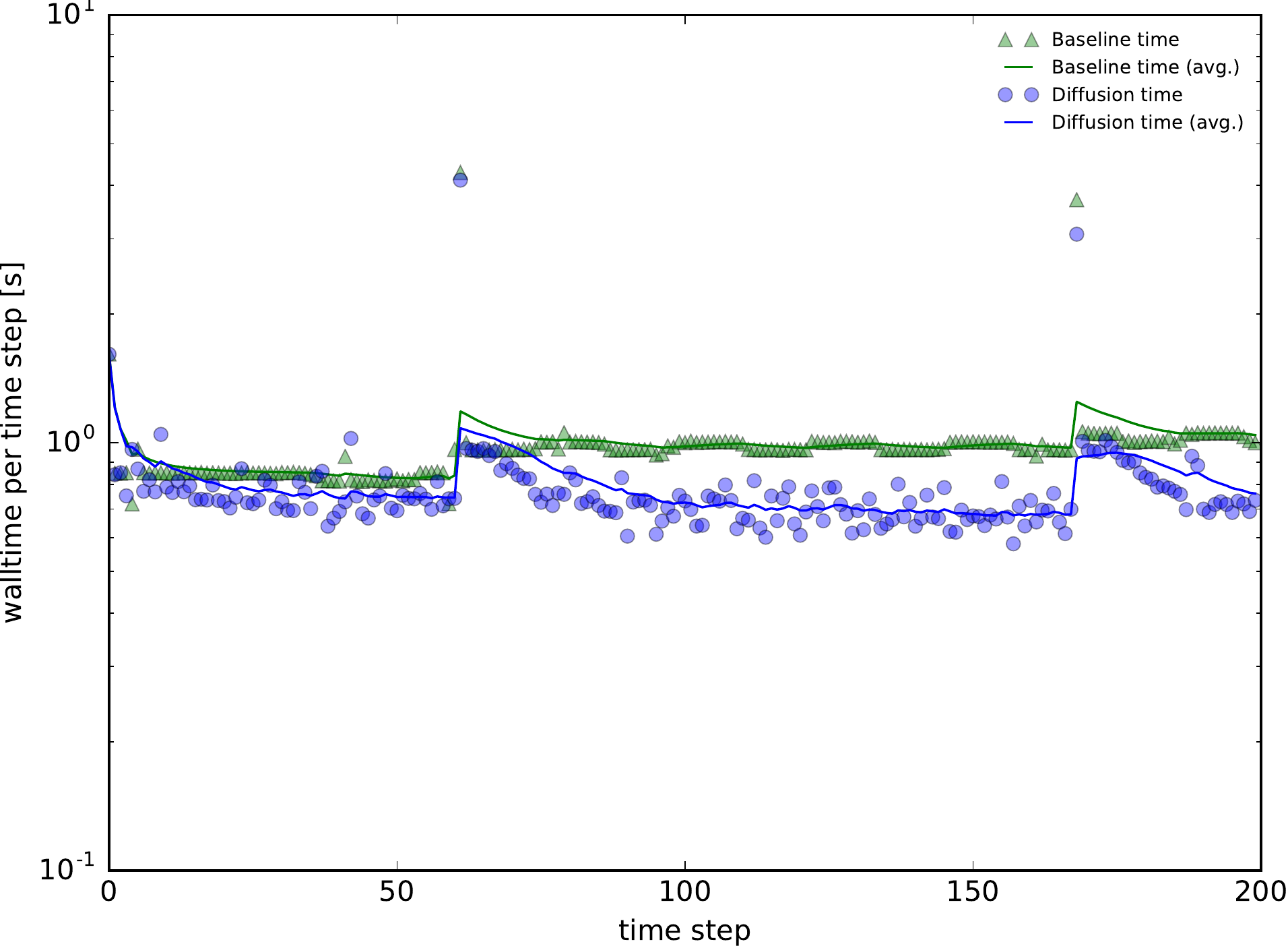}
\caption{\added[id=CCPE1]{Time per timestep for a setup with dynamic AMR on 28 ranks on SuperMUC-NG. }}
\label{fig:dynamic_amr}
\end{figure}

\added[id=CCPE1]{
 Dynamic AMR yields peaks in the runtimes which
 are subsequently damped out by our diffusion (Fig.~\ref{fig:dynamic_amr}).
 If the mesh changes dramatically and thus requires geometric rebalancing, the
 rebalancing cost amplifies the peak yet diminishes the subsequent tail, as it
 directly yields a relative balanced decomposition again.
 Domain repartitioning restarts our offloading yet with a good or even
 optimal initial guess of a partitioning.
 An alternative case of imbalancing results from the temporary
 degradation of node performance as it arises from temporary energy cuts due to overheating,
 hardware failures, co-scheduling or congestion.
 It becomes extreme if the nodes recover quickly again.
 Performance runtime peaks, i.e.~efficiency break-downs, as discussed for the
 traditional balancing challenges here do arise, too, yet cannot be recovered
 and amortised due to task diffusion over the subsequent time steps.
 We thus focus on this last scenario from hereon:
 We (artificially) delay one rank in a 28 rank setup by 1s every ten time steps.
}

\begin{figure}
 \begin{center}
  \includegraphics[width=0.7\textwidth]{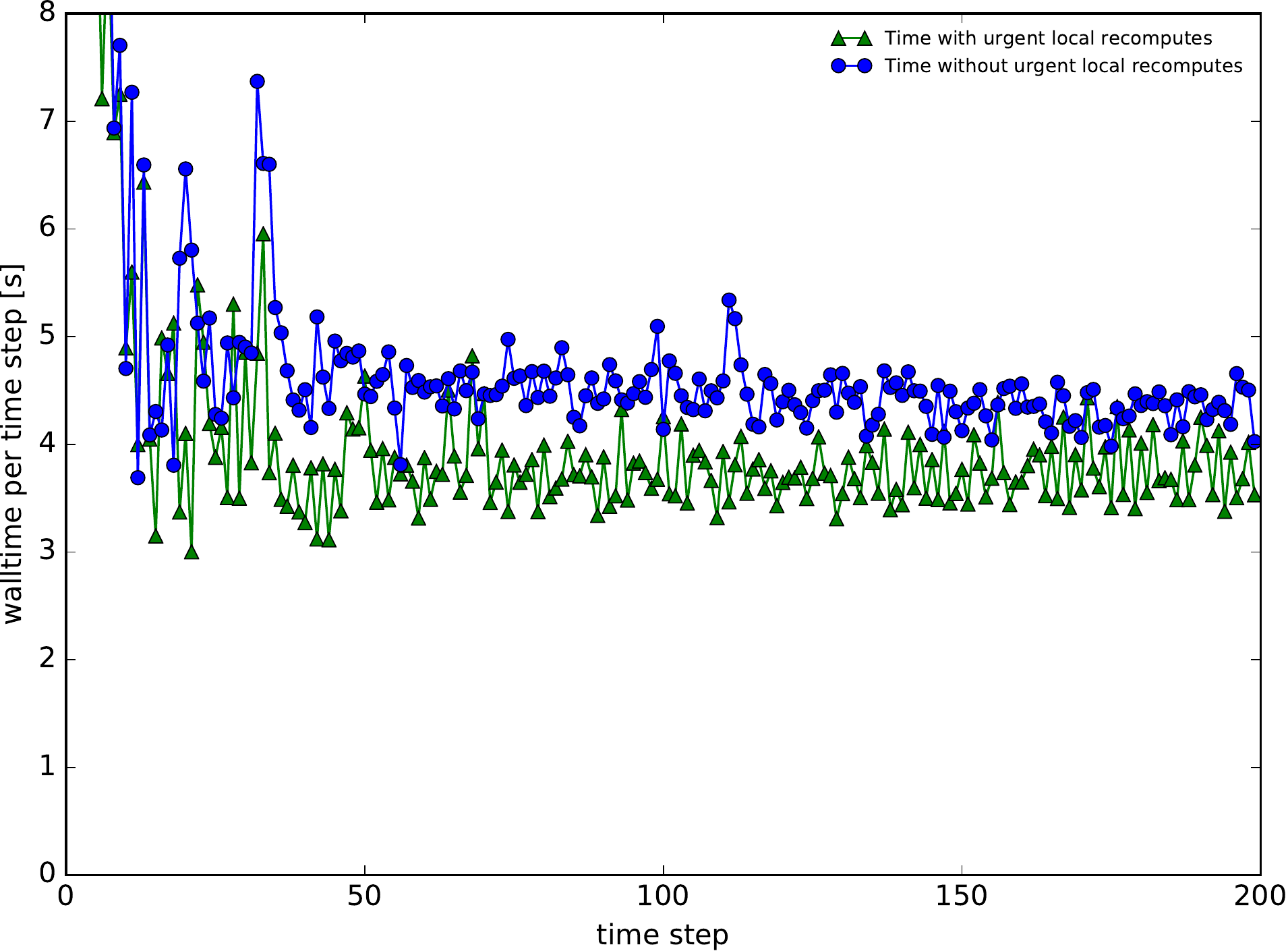}
  \caption{
   \added[id=CCPE1]{
    Runtime per time step for a 28 rank setup, where one rank is delayed by 1s
    every 10 time steps.
    We compare our offloading with urgent recomputes to the offloading without
    them (SuperMUC NG).
   }
   \label{figure:results:urgent-recomputes} 
  }
 \end{center}
\end{figure}
 
\added[id=CCPE1]{
 The peaks every ten time steps make the time per time step flatter
 (Fig.~\ref{figure:results:urgent-recomputes}).
 The peaks are hard to spot, as temporary delays have both immediate
 effects---they delay classic boundary data exchange---as we well as knock-on
 effects due to a delayed delivery of outsourced tasks of the next time step as
 well as an impact on the diffusion metrics.
 Once we enable urgent local recomputes, the fluctuation of runtimes does not 
 reduce dramatically; in particular the load diffusion continues to suffer from
 the strongly changing cost reported.
 However, the major peaks for an unbalanced task distribution are damped out and
 we improve the long-term time-to-solution by roughly 5\%.
 The local urgent recomputes make the task distribution scheme really reactive
 and help to manage dynamically changing setups.
}

%% file: 08d_speedup.tex
\subsection*{Scaling studies}
We continue our evaluation with some scaling studies.
For this, we start with SuperMUC phase 2 and
 use the runtime per time step per degree of freedom on a single
node as baseline.
We normalise against the 28-core single node speed.
Our data span 200 time steps, but we distinguish the runtimes within the first
50 iterations from the measurements within the remaining 150 steps. 
All following setups employ $\omega^\text{(diff)}=1$ and $\omega^\text{(reinf)}=1$.

\input{data/strong-scaling.tex}

Our first set of experiments (Table \ref{results:speedup:strong-scaling}) study
solely setups where we ensure that the geometric load balancing for the regular
grid baseline is close to perfect.
The baseline scaling thus is good, too.
While the reactive diffusion improves upon the regular grid runtimes for the smaller node
choices, its contribution is limited through the strong scaling regime:
If the nodes' workload decreases, we eventually have enough cores available: 
It is cheaper to process tasks locally rather than to give them away---a
decision encoded into our starvation check in Alg.~\ref{algorithm:spawn}.

\input{data/amr-scaling.tex}

With AMR, reactive load balancing robustly 
improves the walltime for all experiments with limited node
counts (Table \ref{results:speedup:amr-scaling}). 
\added[id=CCPE1]{
The improvement is very significant for static AMR.
As we start from a regular grid, partition this grid perfectly, and then add the 
(static) refinement, our diffusion manages to compensate for any illbalancing 
that results from the AMR.
For real-world setups, it might be more convenient to add an additional
rebalancing step once the grid has become stationary, i.e.~to have both an
initial partitioning to facilitate a geometric mesh construction plus a very
good domain decomposition afterwards.
While our diffusion has been designed to act on top of such a load balancing,
the data show that it can also replace the rebalancing step in some scenarios.
For dynamic AMR, the mesh continues to change gradually over time.
The load imbalances are typically small in the beginning but tend to increase in the long term. An example of this behaviour is illustrated in Fig.~\ref{fig:dynamic_amr},
where the dynamic AMR results in two prominent peaks in the time per time step due to the re-meshing. 
There is a slight put persistent increase in time per timestep after each re-meshing step. The diffusion
adapts quickly to this small load imbalance without the need for an expensive 
global repartitioning step. For the other dynamic AMR setups on $20$ nodes in Table \ref{results:speedup:amr-scaling},
the overall observed load imbalance due to dynamic AMR is not large enough to justify a global re-balancing
step. Yet, our reactive load balancing adapts and improves time to solution.}

A major selling point of AMR is its capability to allow codes to scale up
problem sizes in a fine granular way, while work is invested where it pays off
most.
Equidistant global refinement in contrast would make the degrees of freedom and,
hence, the memory footprint explode.
As load balancing for varying grids is challenging, it is here where our
approach helps most. 
It leverages the pressure to re-balance all the time and can compensate for
slight ill-balancing.

\begin{figure}
 \begin{center}
  \includegraphics[width=0.7\textwidth]{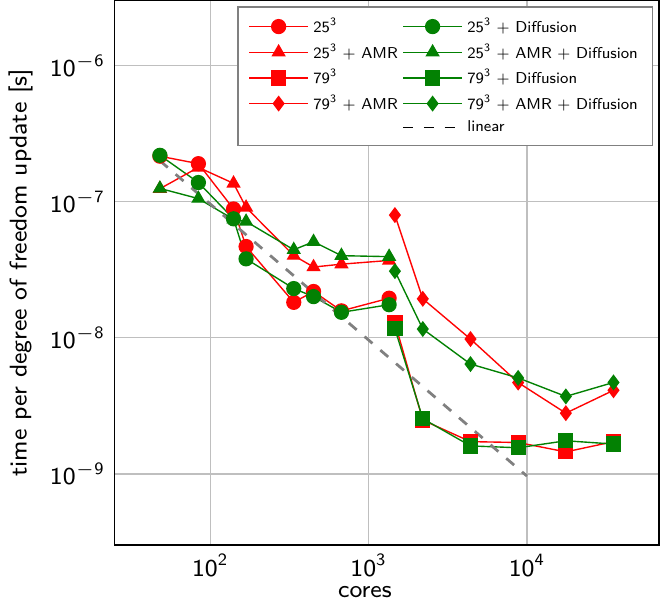}
 \end{center}
 \caption{
  Strong scaling plots for various
  problem sizes with and without AMR on up to 731 ranks (SuperMUC-NG).
  \label{figure:results:strong-scaling-NG}
 }
\end{figure}

 We next benchmark our code systematically on multiple nodes of Super\-MUC-NG on up to 731 ranks
 (Fig.~\ref{figure:results:strong-scaling-NG}).
 Two regular grids of $25\times 25 \times 25$ or $79 \times 79 \times 79$ 
 serve
 as starting point.
 We validated that the chosen geometric load balancing
 approach balances the regular grid setups almost perfectly.
 Indeed, we observe
 reasonable strong scaling behaviour for these regular grid configurations,
 i.e.~runtimes decrease close to linearly with increasing core counts before
 they enter a stagnation regime.
 Our reactive load balancing does not make a real difference for these
 reasonably balanced setups.
 The important observation is, however, that it also does not impose any
 significant runtime penalty. This is due to its totally non-blocking implementation.
 
 We finally allow our adaptivity criterion to add further cells to the 
 regular base grids. 
 For $25^3$ and $79^3$ this yields around $39 \cdot 10^6$
 or $833 \cdot 10^6$ degrees of freedom, respectively.
 The baseline balancing here struggles to yield perfect decompositions.
 Indeed, we observe that the performance curves suffer from some offset,
 while the increased number of degrees of freedom, compared to the regular baseline
 grid, ensures that we scale to slightly more cores.
 Our reactive load balancing manages to narrow this gap between cost per degree
 of freedom in a perfectly balanced world vs.~a world where we have to pay for
 the adaptivity and the resulting illbalancing. However, at large node counts,
 we run into the aforementioned issues (compare Fig.~\ref{figure:results:tasks-per-rank}),
 where the task offloading is limited in the number of tasks that it can offload due to possible 
 starvation of local cores. Indeed, some offloading-related overhead becomes
 visible.

%% file: data/strong-scaling.tex
\begin{table}
  \centering
  \caption{
    Strong scaling speedups.\deleted[id=CCPE1]{for a $25 \times 25 \times 25 $
    baseline grid.}
    We separate time steps 1--25 (top) from 26--200 (bottom).
    \added[id=CCPE1]{For the regular grid experiments, we use a $25 \times 25
    \times 25 $ grid.} 
    AMR \replaced[id=CCPE1]{denotes}{means} that we add one level of AMR to
    \replaced[id=CCPE1]{this regular}{the base} grid.
    The \replaced[id=CCPE1]{data}{reactive load balancing (LB)} columns show by
    which factor the baseline scalability \added[id=CCPE1]{(without task
    offloading)} is improved.
    Entries smaller than 1 denote a slow-down, higher is better.
    \label{results:speedup:strong-scaling}
  }
  \begin{tabular}{r|r|rr}
   Nodes & regular grid & AMR
   \\
   \hline
   2  &  1.14 & 1.24 \\
   4  &  1.10 & 1.33 \\
   7  &  0.90 & 1.05 \\
   14 &  0.92 & 0.90 \\
    \hline
   2  &  0.98 & 2.21 \\
   4  &  1.19 & 1.80 \\
   7  &  0.90 & 1.07 \\
   14 &  0.85 & 0.88
  \end{tabular}
\end{table}


%% file: data/amr-scaling.tex
\begin{table}
  \centering
  \caption{
    Some typical reactive diffusion timings for various numbers of ranks. 
    The mean runtime per time step for 200 time steps is given.
    The label ``dyn'' stands for dynamic
    AMR whereas ``stat'' denotes static AMR. We use a single refinement level for both variants of AMR.
    For dynamic AMR, the number of tasks \#$\mathcal{P}$ is changing over time.
    \label{results:speedup:amr-scaling}
  }
  \begin{tabular}{rrrrr|rr}
   Machine & Ranks & Nodes & \#$\mathcal{P}$ & AMR & Base [t]=s & Diffusion [t]=s \\
   \hline
   Phase 2 & 4    &  1    &   -            & dyn &  18.0 & 14.4 \\ 
   Phase 2 & 7    &  1    &   -            & dyn & 15.3 & 12.7 \\
   NG      & 28   &  2    &   -            & dyn & 4.2 & 3.8 \\
   NG      & 120     &  20   &   -            & dyn &  19.9   &  19.2   \\
   NG      & 480     &  20   &   -            & dyn &   8.9  &   8.5  \\
   Phase 2 & 40   & 20    &   33,201            & stat & 11.2  & 5.1  \\     
   Phase 2 & 140   & 20    &   33,201            & stat & 8.1  & 4.6 \\     
   Phase 2 & 280   & 20    &   33,201            & stat & 5.7 & 3.6 \\     
  \end{tabular}
\end{table}

%% file: 09a_discussion.tex
\section{Discussion}
\label{section:discussion}

%
%
We introduce a very lightweight task migration pattern---lightweight in a sense
that the baseline implementation is hardly changed---which allows us to use time
otherwise spent in MPI waits for actual work.
Many task systems already can exploit MPI waits to process (local) tasks, and
our demonstrator realises this feature, too.
However, we hypothesise that---in almost all cases---such an eager
processing of ready tasks introduces idle time later down the line.
It is thus reasonable to lighweightly ``fill up'' wait time with remote
tasks from ranks that are overbooked.
As our scheme migrates tasks non-persistently, this feature
is particularly appealing for machines that suffer from speed fluctuations and
for simulations where the load balancing is constrained due to the main memory
available or load balancing overheads.
\added[id=CCPE1]{%
Different to other approaches such as \cite{Garcia:09:LeWI} that translate the
concept of task stealing into a distributed memory world, our scheme furthermore
proactively outsources tasks, i.e.~we try to have the migrated tasks in place before the actual wait occurs.
Otherwise, inter-node latency would become a challenge.
}

%
%
There are natural shortcomings of the present approach.
First, our algorithms focus on ready tasks only.
Tasks that have dependencies \cite{Bremer:18:SemiStaticLB,Schaller:16:Swift} are
not supported.
On the long term, it is interesting 
 to
migrate whole task assemblies if a task set as a whole
requires less data per computation to exchange than
its individual tasks.
\added[id=CCPE1]{%
 Migrating task subgraphs also helps in situations where the number of ready
 tasks alone is too small or just big enough to keep the local cores busy.
 That is, it helps whenever migrating load of ready tasks would compromise the
 local occupation.} %
Second, our algorithms are not yet memory aware. 
There is no quota on the maximum number of
stolen tasks hosted by a victim.
Victims consequently might exceed their memory.
Memory consumption could be another blacklisting criterion.
Third, we have chosen several ``magic'' parameters for our experiments.
While they yield meaningful results, we can not claim that they are optimal.
Autotuning 
here might improve the code's performance.~\cite{Eckhardt:15:SPHCompression}
Finally, 
it might be reasonable to take the network topology 
as well as the logical rank topology
into account when a rank selects its victim.
Rather than choosing the most underbooked rank globally, we could
offload tasks to nearby ranks. %
This constrains the task migration but avoids that offloading adds more edges to the logical MPI
communication graph that many codes tailor
towards a network architecture.

The weakest point we see in terms of methodology is the lack of an appropriate
notion of criticalness.
Our experiments run into situations where victim ranks are given too many remote
tasks.
This delays their actual delivery of information such as boundary data and
eventually slows down critical ranks further.
They however do not recognise this as they are not waiting for an outsourced
task.
Such complex causal dependencies can not be tracked by our current notion of an
emergency.
We track overloading in a compute sense, but lack a detector for overloading in
a bandwidth or MPI overhead (too many pending non-blocking messages) sense.

We have extensively invested into a scheme which ensures that reasonable
progress is made on the asynchronous MPI transfers
without sacrificing a thread \cite{Hoefler:08:SacrificeThread}.
We use aggressive polling.
Yet, this cannot detect congestion.
While the prioritisation of MPI messages might mitigate this problem to some degree, 
we would appreciate if there were an  
MPI monitoring, i.e.,~online performance analysis
that can tell the application if the MPI subsystem enters a critical state.
This could be realised via software
\cite{Mao:14:WaitStateProfiler} supervising the machine state.
Alternatively, ``intelligent'' communication devices alike the
SmartNIC technology could host the monitoring.

On the shared memory side, it remains open to which degree our choice of 
TBB as tasking base with manual tweaking of features alike prioritisation affects
the performance results.
All proposed software building blocks currently are extracted into a
standalone software package such that they can be used more easily with other
codes~\cite{Chameleon:19}.
As part of this roll out, we also explore the integration into OpenMP.
On the long term an abstraction over various tasking paradigms
\cite{Alomairy:19:TaskAPI} however might become necessary, such that we can
systematically study the interplay of tasking approach and our balancing.

Our lightweight task migration realises push semantics:
Oversubscribed ranks deploy work to other ranks.
This approach differs to strategies where ranks know their task workload prior
to the computation---though they might permanently renegotiate, i.e.~balance such
responsibilities---or codes with pull semantics, where ranks ``grab''
tasks from a (distributed) repository \cite{Meng:2010:Uintah}.
While all paradigms might yield comparable data distribution graphs, our
code migrates tasks only temporarily, i.e.~sends results back.
Our induced data flow graph is cyclic.
It is thus lightweight as it does not redistribute data permanently.
It is not lightweight by means of data moves, as every temporary task migration
relies on a send forth and a send back.

%% file: 09b_conclusion.tex
\section{Outlook}
\label{section:outlook}

%
%
The exact interplay 
of our scheme
with various dynamic load balancing schemes or more
sophisticated numerics is beyond scope
for the present paper.
We do however expect that our approach has beneficial knock-on effects:
If load balancing is semi-static \cite{Bremer:18:SemiStaticLB}, i.e.,~rebalanced
only every $k$ steps, we may assume that our approach allows us to migrate work
less frequently (similar to \cite{Samfass:18:ReactiveSamoa}).
This reduces AMR overhead.
If load is balanced continuously in a diffusive style, we may assume that the
diffusion rate, i.e.,~the amount of data migration per step, can be chosen
smaller with our approach.
This reduces bandwidth requirements.
On accelerator-driven machines, where bandwidth and local memory are
notoriously short, we may assume that our approach offers an
alternative to the difficult heterogeneous scheduling
\cite{Sundar:15:Enclave}.
Our approach would make each accelerator a designated victim
and thus hide the complexity of persistent data migration to balance load between accelerators.

On the numerics side, we will investigate non-linear equation systems in
the ADER-DG context. 
Such schemes require iterative Picard or Newton solves per $\mathcal{P}$ task
\cite{zanotti:2015}. This renders the cost per $\mathcal{P}$ evaluation very hard to predict.
ADER-DG is often contrasted with standard Runge-Kutta (RK) methods. 
Indeed, our ideas should apply to RK as well, though their lack of a space-time
evaluation might imply that the evolution of the cells is cheaper. 
In return, we might get away with a smaller memory footprint.
This renders RK another interesting numerical scheme to study.
ADER-DG's attractiveness is its inherent fit to adaptive, local time stepping.
Again, such a time stepping renders the workload prediction very difficult.
It thus should benefit from our approach.
Finally, we plan, on the long term, to study the interplay of our Eulerian
mindset with Lagrangian techniques 
(see \cite{Dubey:11:LagrangianInEulerianMeshes} or \cite{Weinzierl:15:PIC}, 
 the latter also working on tree-structured adaptive Cartesian grids).
While the load inhomogeneity resulting from these couplings is obvious, it is
notably the fact that such setups have to balance both memory and compute load
rigorously which makes it interesting for our approach.

%% file: 99_acknowledgements.tex

\section*{Acknowledgements}

This work has received funding from the European Union's Horizon 2020
research and innovation programme under grant agreement No 671698 (ExaHyPE).
\added[id=CCPE1]{%
 Tobias has received additional funding around ExaHyPE through 
 EPSRC's Excalibur programme under grant number EP/V00154X/1
 (ExaClaw).%
}
 We also acknowledge support and computing resources provided by the Leibniz Supecomputing Centre (grant no\ pr48ma).
 Special thanks are due to all members of the \mbox{ExaHyPE} consortium who made this research
 possible -- in particular to Leonhard Rannabauer for his work on the elastic wave equation solver and the LOH.1 setup.
 All underlying software is open source \cite{Software:ExaHyPE}.
